\documentclass[12pt]{iopart}
\usepackage{amssymb}
\usepackage{iopams}
\usepackage{amsmath,amsfonts}
\usepackage[dvipsnames]{color}
\usepackage{graphicx}% Include figure files
\usepackage{dcolumn}% Align table columns on decimal point
\usepackage{bm}% bold math
\usepackage{multirow}
\usepackage{url}
\usepackage{diagbox,booktabs}
\usepackage[font=small,skip=5pt]{caption}
\usepackage{subcaption}
\usepackage[usenames,dvipsnames,svgnames,table]{xcolor}
\usepackage{CJK}
\usepackage{indentfirst}
\usepackage{epstopdf}
\usepackage[ruled, lined]{algorithm2e}
\usepackage{setspace}
\usepackage{enumerate}

\usepackage{xcolor}

\graphicspath{{Figures/}}

\def\lsim{\mathrel{\rlap{\lower4pt\hbox{\hskip1pt$\sim$}}
    \raise1pt\hbox{$<$}}}

\begin{document}

%\title[GPU-accelerated CBC search]{GPU-accelerated low-latency real-time searches for gravitational waves from compact binary coalescence}
\title[GPU-accelerated CBC search]{Acceleration of low-latency gravitational wave searches using \textit{Maxwell}-microarchitecture GPUs}

% repeat the \author .. \affiliation  etc. as needed
% \email, \thanks, \homepage, \altaffiliation all apply to the current
% author. Explanatory text should go in the []'s, actual e-mail
% address or url should go in the {}'s for \email and \homepage.
% Please use the appropriate macro foreach each type of information

% \affiliation command applies to all authors since the last
% \affiliation command. The \affiliation command should follow the
% other information

% \affiliation can be followed by \email, \homepage, \thanks as well.

\author{Xiangyu Guo$^1$, Qi Chu$^2$\footnote{Contributed equally to this work with Xiangyu Guo, Corresponding Author's Email: qi.chu@uwa.edu.au}
, Shin Kee Chung$^2$, Zhihui Du$^1$\footnote{Corresponding Author's Email: duzh@tsinghua.edu.cn} and Linqing Wen $^2$\footnote{Corresponding Author's Email: linqing.wen@uwa.edu.au}}

\address{$^1$
 Tsinghua National Laboratory for Information Science and Technology, Department of Computer Science and Technology, Tsinghua University, Beijing, 100084, China
}
\address{$^2$
 School of Physics, The University of Western Australia, M468, 35 Stirling Hwy, Crawley, WA 6009, Australia
}

%Collaboration name if desired (requires use of superscriptaddress
%option in \documentclass). \noaffiliation is required (may also be
%used with the \author command).
%\collaboration can be followed by \email, \homepage, \thanks as well.
%\collaboration{}
%\noaffiliation

\date{\today}

\begin{abstract}
Low-latency detections of gravitational waves (GWs) are crucial to enable prompt follow-up observations to astrophysical transients by conventional telescopes. We have developed a low-latency pipeline using a technique called Summed Parallel Infinite Impulse Response (SPIIR) filtering, realized by a Graphic Processing Unit (GPU). In this paper, we exploit the new \textit{Maxwell} memory access architecture in NVIDIA GPUs, namely the read-only data cache, warp-shuffle, and cross-warp atomic techniques. We report a 3-fold speed-up over our previous implementation of this filtering technique. To tackle SPIIR with relatively few filters, we develop a new GPU thread configuration with a nearly 10-fold speedup. In addition, we implement a multi-rate scheme of SPIIR filtering using Maxwell GPUs. We achieve more than 100-fold speed-up over a single core CPU for the multi-rate filtering scheme. This results in an overall of 21-fold CPU usage reduction for the entire SPIIR pipeline.% by exploiting the new GPU-memory access and exchange mechanism of the recent GPU microarchitecture --- \textit{Maxwell}. We incorporated the GPU accelerated multi-rate scheme in the low-latency SPIIR detection pipeline. The CPU usage of the pipeline is reduced by 21-fold. %The details of our GPU acceleration and performance is presented.
%Our new SPIIR acceleration achieve 10-fold speedups compared to the previous GPU acceleration implementation thanks to the ever improving GPU technology. %We show that our new method, implemented and tested on the latest Maxwell microarchitecture NVIDIA GTX 980 GPU, can achieve up to 125-fold speedups compared with a single Intel i7 3770 CPU.   
%We show that the main computational part of the low-latency SPIIR coincident search pipeline is the multi-rate filtering scheme. When applying the GPU acceleration of this scheme on the search pipeline, a $21\times$ improvement in reduction of CPU resource usage is achieved.

%\R{(shinkee: Need to reconsider whether we want to put such detail in abstract.)}
%The difference of the signal-to-noise ratio results between the GPU-powered pipeline and the CPU pipeline on the injection is within $0.00006\%$ with the 10 runs.

%Furthermore,  the proposed adaptive template mapping mechanism can achieve the GPU resource utilization rate to nearly 100\%. 
%With the new method, the SPIIR filtering is now capable of processing up to 23000 templates per GPU in real time.  This indicates that a comprehensive search for gravitational waves from coalescences of binary neutron stars and black holes in real time is possible with a fairly small number of GPUs for the advanced LIGO detectors.

\end{abstract}

% insert suggested PACS numbers in braces on next line
 %Gravitational wave detectors and experiments- Gravitational radiation detectors; mass spectrometers; and other instrumentation and techniques- Observation and data reduction techniques; computer modeling and simulation - gravitational radiation-Binary and multiple stars
%\keywords{gravitational waves --- binaries:close}
% insert suggested keywords - APS authors don't need to do this
%\keywords{}

%\maketitle must follow title, authors, abstract, \pacs, and \keywords
%\maketitle

% body of paper here - Use proper section commands
% References should be done using the \cite, \ref, and \label commands
\section{Introduction}
%The first GW event by LIGO twin detector, GW150914, is initially alarmed from a low-latency search pipeline. It had been through careful validation and the source is confirmed as a black hole binary~\cite{detectionpaper}. 

We are entering an exciting time of gravitational wave (GW) astronomy. The first GW signal is detected in September 2015 by the Laser Interferometric Gravitational-Wave Observatory (LIGO)~\cite{first_discovery}. This opens up a new window to multi messenger astronomy with unprecedented power of discovery, in probes of some of the most enigmatic transients in the sky for their emissions in both the electro-magnetic and gravitational wave spectrum, e.g., short or long gamma-ray bursts produced in binary coalescence and core-collapse supernovae~\cite{sathya09physics,van2014origin}. In particular, low-latency detection and localization of GW sources are gaining priority in order to enable prompt electromagnetic (EM) follow-up observations of GW sources. Capture of transient EM events triggered by low-latency GW events will bring breakthroughs in understanding the mechanism of sources~\cite{chichi_2016MN}.

The optimal method to search for compact binary coalescence (CBC) GW sources is by correlating the expected waveform templates with data, known as the matched filtering method~\cite{matchedfiltering,bruce05}. An efficient way to compute the correlation is by the fast Fourier transform (FFT) technique, especially so when the array size is large. Low latency searches, however, require real-time processing of relatively short data-segments, limiting any gain obtained in FFT-based correlations. Existing efforts of low-latency GW detection from binary coalescence sources by LIGO-Virgo community employ sophisticated strategies to realize low-latency with manageable computation, notably a Multi-Band Template Analysis (MBTA) \cite{MBTA} method utilizing the two-band frequency-domain Finite-Impulse-Response (FIR) filters to represent a given template for filtering , the Low-Latency Online Inspiral Detector (LLOID, a.k.a GstLAL for the software name) \cite{LLOID} method utilizing multi-band FIR filters and the Summed Parallel Infinite Impulse Response (SPIIR) \cite{jing, shaun, shaun2} method utilizing infinite impulse response (IIR) technique in the time domain which is expected to have zero latency. These three techniques evolve into automated software programs, known as \textit{pipelines}, which additionally allow input from multiple detectors and employ template-based statistical tests to veto transient noises and produce GW alerts to the community. As of this writing, these three pipelines have achieve a medium latency of less than one minute. This paper is focused on the SPIIR method. Compared to the FFT technique, it is expected to be more efficient at latencies lower than tens of seconds for advanced detectors~\cite{jing}.

%The tradeoff is that low-latency search requires that the data segment processed by the FFT method to be as small as possible, while in consideration of computation efficiency, it is best for the FFT technique to take longer segment of data. 
% This technique involves transformation of a segment of data from the time domain to the frequency domain.

% Other than the filtering process, the three pipelines fullfil a set of functionalities that can take data from multiple detectors and produces GW alerts as output. We refer \emph{latency} as the delay from the time a GW signal actually occurs to the time a search pipeline sends out an alert for this GW signal. 
%In the mean time, the large template space of binary coalescences to search for places a computational challenge on search pipelines. The aim of this paper is to deal with the challenge for the SPIIR pipeline using high-performance computing platform, the GPUs. The acceleration technique presented in this paper may provide insights, if not be used directly, to acceleration of other search pipelines. 

%GPU is first introduced in 1999~\cite{mcclanahan2010history} to address the computation demands of graphics rendering. It is then adapted to provide solution for general purpose problems as general purpose graphics processing unit (GPGPU) since 2006. GPU acceleration for GW search is first introduced by Shinkee Chung~\cite{shinkee}. The implementation has acheived 16-fold speed-up and is later incorporated into the LIGO algorithm library~\cite{lal}.

Our previous work of GPU accelerated SPIIR filtering method uses NVIDIA's \textit{Fermi} GPUs with a speedup a factor in the order of 50-fold over a single core Intel i7 CPU~\cite{liu2012gpu}. However, that GPU optimization targeted the SPIIR filtering with number of SPIIR filters in the range of 128 to 256. While in the multi-rate filtering scheme, the total filters are split to be applied to data at different sampling rates, that the number of filters can be as small as a few in some sampling rates. In this paper, we extend the GPU optimization of SPIIR filtering for various number of filters and explore the features of the recent \textit{Maxwell} GPUs. Our optimization here achieves 3-fold improvement over the previous targeted range and achieves up to 10-fold speedup beyond the range, compared to the previous work. Compared to the SPIIR filtering on a single-core CPU, our GPU acceleration is now 60 to 125 fold faster depending on the number of filters to be applied.
%a GPU resource utilization to the extent of approximately 100 percent and up to 125-fold speedups compared with the CPU counterpart, 

In addition to the SPIIR filtering, we have extended the GPU acceleration to other components of the SPIIR pipeline. Another bottleneck of the SPIIR pipeline in terms of computational efficiency is the sampling-rate alterations of multi-rate filtering. In multi-rate filtering, a set of filters is applied to data at different sampling rates, the results of which are combined to the initial rate. Our acceleration of the multi-rate filtering scheme realizes a $100$-fold improvement in CPU resource reduction and hence a $21$ fold reduction in CPU resource for the entire SPIIR pipeline. The filtering process in this paper uses single-precision floating-point number format, the result of which has ignorable difference with the results using double-precision format. 
%whitened data and whitened templates, that are within the description range of  The acceleration in this paper is based on the It is a widely used strategy to reduce computational cost in GW signal processing. 

The structure of this paper is as follows: In Sec.~\ref{sec:multi-rate} we review the mathematical representation of the SPIIR method and present the multi-rate filtering scheme for our GPU implementation. In Sec.~\ref{sec:optimization} we present the technical details of our new GPU optimization on SPIIR filtering exploiting the newest features from Maxwell GPUs and GPU optimization technique on sampling-rate alterations. In Sec.~\ref{sec:er} we present the performance results of our GPU accelerated SPIIR filtering and the multi-rate filtering scheme. In Sec.~\ref{sec:pipeline}, we present the performance of our GPU accelerated low-latency SPIIR detection pipeline. The conclusion and future work will be given in Sec.\ref{sec:con}.

\section{Multi-rate SPIIR filtering}
%\label{sec:spiir}
\label{sec:multi-rate}

%\subsection{Infinite Impulse Response Filter}
\subsection{SPIIR method}
Matched filtering is known to be the optimal detection method in deep searches for signals in stationary and Gaussian noise. Here, we consider a CBC waveform template in the time domain $h(t)$. The optimal detection output, known as the signal-to-noise ratio (SNR), is given by:

\begin{equation}
\label{equ:zcontinue}
z(t) = \int_{-\infty}^{t}{x(t')h(t' - t)dt'},
\end{equation} 
i.e.~the cross-correlation between whitened waveform template $h(t)$ and the whitened detector strain $x(t)$, which is given by:

%We define $x$ as the over-whitened detector strain data,

\begin{equation}
x_{t} = \int_{-\infty}^{\infty}{\frac{\tilde{s}(f)}{\sqrt{S_{n}{(f)}}}e^{2{\pi}ift}df},
\end{equation}
where $\tilde{s}(f)$ is the Fourier transform of detector data s($t$) and $S_{n}(f)$ is the one-sided noise spectral density of a detector defined through the expectation $E$:
\begin{equation}
\label{eq:psd_def}
E\left(\tilde{n}(f)\tilde{n}^*(f')\right) = \frac{1}{2}S_n(f)\delta(f-f').
\end{equation}

By sampling at discrete instances $t=k\delta t$ at a sampling rate $1/\delta t$ $(k =0,1,\cdot)$, Eq.~\ref{equ:zcontinue} can be rewritten in a discrete form,
\begin{equation}
z_{k} = \sum_{j=-\infty}^{k}{x_{j}h_{j-k}{\Delta{t}}}.
\end{equation}

CBC searches usually adopt the matched filter $h^*(-t)$, that takes Eq.~\ref{equ:zcontinue} into a convolution integral. Quite different from the common FIR method, SPIIR method utilizes IIR filters to reconstruct the matched filter. A first-order IIR filter bears a simple form shown in Eq.~\ref{equ:first-order1} and Eq.~\ref{equ:first-order2}:  %IIR filters are digital filters with infinite impulse response. The difference between FIR filters and IIR filters is that the history of IIR filtering outputs will roll off as a part to form the filtering result of this round.  A example of a first-order IIR filter is shown in eqation(\ref{equ:spiir}). Usually much less IIR filters than FIR filters are needed for the same functionality of digital filtering. the time-reversed counterpart of a waveform template, as the FIR filters to process data

\begin{equation}
\label{equ:first-order1}
y_{k} = a_{1}y_{k-1} + b_{0}x_{k},
\end{equation}
where $y_{k}$ is the filter output at time step $k$ ($t_{k} = k\Delta{t}$), $x_{k}$ is the filter input, and $a_1$ and $b_0$  are complex coefficients. A solution to this first-order linear inhomogeneous difference equation is:

\begin{equation}
\label{equ:first-order2}
y_{k} = \sum_{j=-\infty}^{k}{x_{j}b_{0}a_{1}^{k-j}}.
\end{equation}

For a target as complex as a CBC signal, a group of first-order IIR filters are developed with each filter representing a small segment of the matched filter~\cite{jing, shaun2}. The number of IIR filters that are needed to reconstruct a highly accurate matched filter varies from a dozen to several hundreds, depending on the complexity of the waveform and the the limit of the detection band. The filter construction procedure can be found in~\cite{jing,shaun2}. Here, we express the output of a SPIIR filter as:
%In practice, digital IIR filter design is a more complex process than the FIR design. The IIR filter coefficients are usually obtained by applying well known methods such as bi-linear transform or impulse invariance to a pre-constructed equivalent analog filter.  For IIR filters developed to search for CBC GW signals, a pre-constructed CBC waveform is given in prior to develop IIR filters~\cite{shaun2}. One IIR filter is not enough to filter a given waveform as complex as a CBC waveform. Therefore parallel IIR filters are developed. The outputs of parallel IIR filters are summed together and the overall result is the summation of the all participating transfer functions.

%For the simplest single-pole IIR filter, the difference equation of this filter is
%\subsection{Summed Parallel IIR filtering}

%The filter $h$ is commonly redefined to be a complex signal waveform for searching over unknown phase, following the work of Luan $et$ $ al$ and Hooper $et$ $ al$ \cite{jing, shaun2}, the simplest IIR filter for the CBC signal at $k$th time step can be represented using equation (\ref{equ:spiir}), where $x_{k}$ is the discretized detector strain data. To recover the optimal matched filtering output for the entire CBC signal, a series of IIR filters, which stands for the time evolution of frequency and amplitude of the signal, are used,

\begin{equation}
\label{iireq0}
y_{k, l} = a_{1, l}y_{k-1, l} + b_{0, l}x_{k-d_{l}},
\end{equation}
where $y_{k,l}$ is the output for the $l$th filter and $d_{l}$ is the time-delay for this filter. The discrete form of SNR output from a group of SPIIR filters is given by:%The linear summation of the output of all filters is the cross correlation of the data $x$ and the approximate waveform,

\begin{equation}
\label{iireq1}
z_{k} \simeq 2\Delta{t}\sum_{l}{y_{k, l}}.
\end{equation}

%The SNR is the absolute value of the SPIIR output $z_{k}$ divided by a normalization constant.

\subsection{Multi-rate implementation of SPIIR filtering} \label{sec:multi_design}

%FIXME: a figure here

\begin{figure}[htbp]
	\centering
	\includegraphics[width=10.0cm]{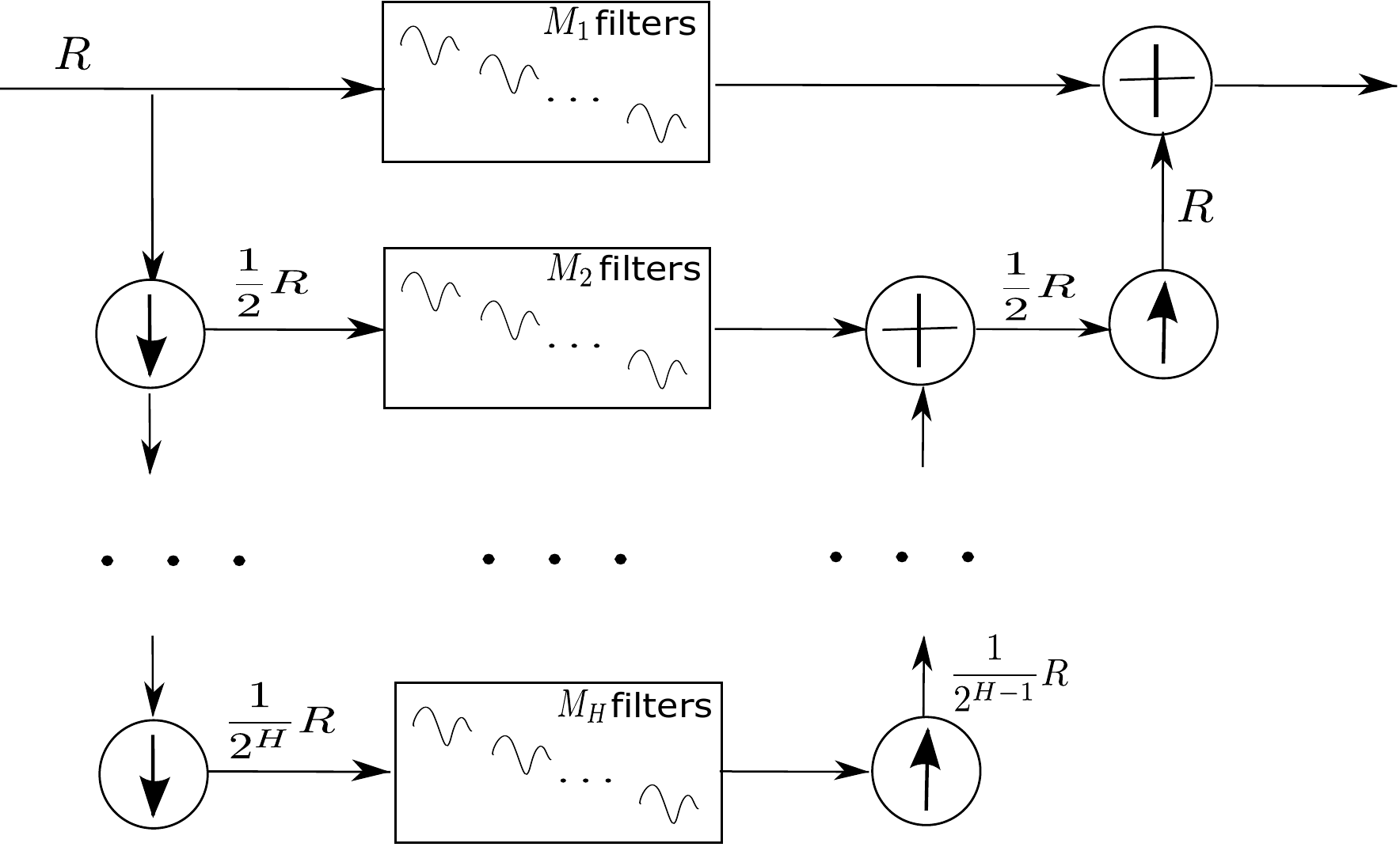}
	\caption{Schematic diagram of multi-rate implementation of SPIIR filtering scheme. The input and output sampling rates are $R$ Hz. $M_1, M_2,...,M_H$ represent the number of filters to be applied to the corresponding rates.}
	\label{fig:multi}
\end{figure}

Our implementation of the multi-rate filtering scheme for GPU acceleration is shown in Fig.~\ref{fig:multi}. For implementation conveniency, the data is downsampled by a factor of 2 in succession. Each sample rate stream is filtered using the corresponding SPIIR filters. The filtering output of the lowerest rate will be upsampled and added to the filtering output of next highest rate. This process will repeat several times until reaching the initial sampling rate. 

Which sampling rate should a given SPIIR filter work on is essentially determined by the $a_{1,l}$ coefficient of the filter. The $a_{1,l}$ coefficient determines the upper bound of the frequency band of the filter, and thus the Nyquist rate~\cite{nyquistrate} for the filter to function. We round the Nyquist rate of the filter to the nearest biggest available rate for the filter to work on.

%Sampling-rate alteration, also known as resampling, is a basic tool in digital signal processing and has seen extensive applications. The theory foundation is Shannon's sampling theorem that signal can be uniquely and exactly reconstructed at any time from its samples by interpolation assuming the frequency band of signal is limited to half the sampling rate.

\begin{equation}
x(m T'_s) = \sum\limits_{n=-(K-1)}^{n=K} x(nT_s) \mathcal{K}(m T'_s - nT_s)
\label{kaiser_window_filtering}
\end{equation}

To avoid the known problems of spectral leakage caused by the squared window, we adopt the popular Kaiser-window-tapered low-pass filter implemented in the open-source \textit{gstreamer} library~\footnote{\textit{gstreamer} library: http://gstreamer.freedesktop.org/} to perform the interpolation for resampling. The resampling formula is shown in Eq.~\ref{kaiser_window_filtering} where $x$ is the data, and $\mathcal{K}$ represents the Kaiser-window-tapered low-pass filter. $m$ and $n$ are discrete sampling points. $F_s=1/T_s$ is the original sampling rate, $F_{s'}=1/T'_s$ is the resampled rate. $x$ is assumed to be bandlimited to $\pm F_s/2$. $K$ is the length of the filter. We call the Kaiser-window-tapered low-pass filter `Kaiser filter' in the rest of the paper.

There is a single parameter that controls the quality, measured by the stop-band attenuation, of Kaiser filter. The \textit{gstreamer} library provides 11 Kaiser filters with stop-band attenuation up to 100 units of decibels (dB). To avoid the band aliasing of downsampling, we choose the Kaiser filter with stop-band attenuation of $\sim100$ dB. We choose the Kaiser filter with stop-band attenuation of $\sim60$ dB for upsampling that we found work most efficiently while maintaing signal recovery quality in practice.

The expected computational efficiency of our multi-rate scheme can be estimated as follows. We denote the number of total SPIIR filters of a given template as $M$; the full-rate we are considering as $R$; and the number of search templates as $N$. According to Eq.~\ref{iireq0} and Eq.~\ref{iireq1}, filtering on one data point requires 12 floating operations. Thus the total floating point operations per second (flops) for SPIIR filtering at full-rate $R$ is $12NMR$. The $100$ dB downsampling Kaiser filter has 384 steps in \textit{gstreamer} and the $60$ dB upsampling Kaiser filter has 32 steps. The resampling and summation of filtering result at different sample rates are negligible. If half of the filters can be applied to sub-rate $\displaystyle \frac{1}{2}R$, the cost will reduce by $25\%$. A factor of a few savings on the computation cost is expected if more filters can be applied to even lower rates..%The cost of combination of two results is $2NR$ at rate $R$. There is a factor of few on computation reduction when comparing the savings from SPIIR filtering and added cost of resampling and combination.

\section{Optimization of Multi-rate SPIIR filtering on Maxwell GPUs} \label{sec:optimization}

There are several interfaces to use GPU for general purpose problem, including NVIDIA's Compute Unified Device Architecture (CUDA) \cite{nvidia2011nvidia} language, Khronos Group's Open Computing Language (OpenCL)~\cite{khronos2008opencl}, Microsoft's C++ Accelerated Massive Parallelism (C++ AMP)~\cite{gregory2012c++}. We use the popular CUDA language for our GPU acceleration.

%Though there are CUDA software libraries available which can be used with minimal knowledge of GPU hardware, to achieve high-performance GPU acceleration, an understanding of the relation of GPU hardware and CUDA semantic is required. 

There are several elements involved in any GPU acceleration, here using CUDA, namely the mapping of algorithm functions to CUDA threads and blocks, and the mapping of data to GPU hierarchical memory architecture. We focus on these optimization strategies accordingly (Sec.~\ref{sec:template_mapping}, Sec.~\ref{sec:data_access}, Sec.~\ref{sec:resampling_addition}). In this process, we set out to exploit the advanced GPU memory exchange mechanism of \textit{Maxwell} GPUs, namely the warp-shuffle and atomic operation techniques (sec~\ref{sec:synchronization}).

%exploited the implicit synchronizations in a CUDA warp to remove the overhead cost of synchronization operations, and inexpensive atomic operations to remove additional shared memory accesses and simplify our implementation

We provide a general explanation of the relation of a GPU hardware and the CUDA semantics for reference here. A GPU chip consists of several Streaming Multiprocessors (SMs). The Maxwell GPU features improved SM architecture renamed as SMM. One SMM has many processing cores upon which one is capable of create, managing, scheduling, and executing CUDA threads. CUDA threads are executed in groups of 32, which are named "warps". A group of CUDA warps aggregate to a CUDA block. While one block is limited to one SMM, one SMM can have several blocks running concurrently. A GPU has a hierarchy of memory spanning a range of access speeds and storage sizes. The choice of memory to use may greatly affect the overall GPU performance.
%This feature aims to maximize its functional units utilization. 
%In hardware level, CUDA employs a SIMT (Single-Instruction, Multiple-Threads) architecture.
%, which is similar to SIMD (Single-Instruction, Multiple-Data) vector organizations in that a single instruction controls multiple processing elements. 

\subsection{Optimization of SPIIR filtering in single-rate}
\subsubsection{Optimization for varying number of filters} \label{sec:template_mapping} 
%The long latency of some instructions can be completely "hidden" by executing other ready warps in the same block on a SMM. 

%As shown in Sec.~\ref{sec:multi-rate}, in the SPIIR method, a template is represented by a number IIR filters where each filter can only respond to a narrow range of input frequency. 

\begin{figure}[htbp]
	\centering
	\includegraphics[width=6.0cm]{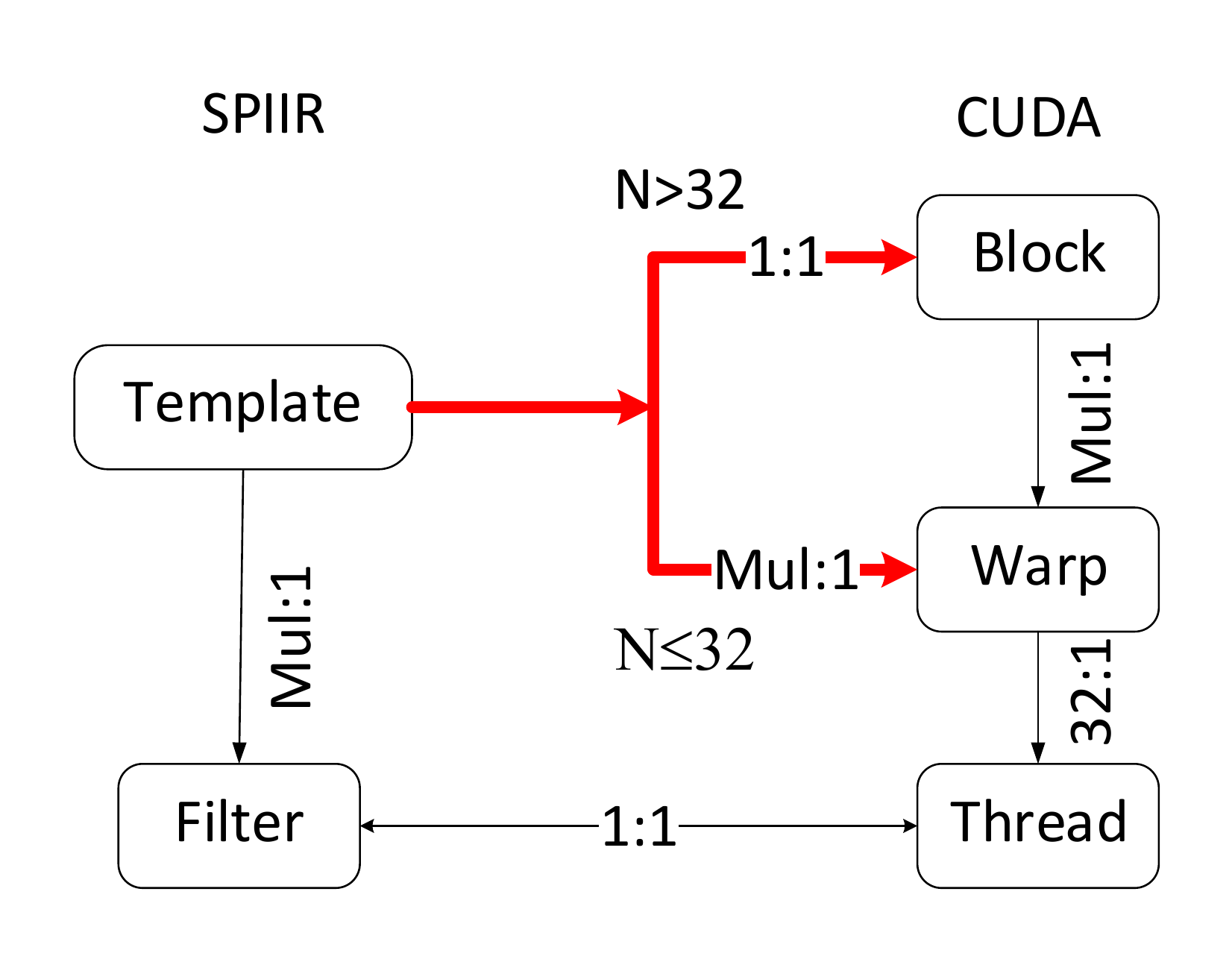}
	\caption{Schematic of the SPIIR template-filters hierarchy mapped onto the CUDA block-warp-threads hierarchy. `Mul' means `Multiple' in this figure. $N$ is the number of filters of any given SPIIR template.}
	\label{fig:mapping}
\end{figure}

Our previous GPU optimization~\cite{liu2012gpu} considers SPIIR filtering with the number of filters a few hundred. It is not highly-optimized toward filtering with small number of filters, which is likely the case in the multi-rate filtering scheme. Here, we develop a CUDA kernel for templates with $\le 32$ filters, which we call warp-based kernel, where the filters in one or multiple templates are mapped to one CUDA warp. This allows us to not only avoid branch execution paths in one warp but also significantly reduce the number of idle threads. We keep the block-based CUDA kernel, presented in the previous work, for templates with $>32$ filters where each template is mapped to one CUDA block with multiple CUDA warps. Fig. \ref{fig:mapping} shows the actual mapping between IIR filters for one template and GPU threads for one block regarding the two kinds of configurations.

%We use different CUDA configuration for different sizes of templates to ensure high occupancy for smaller templates.
%. One is a warp-based CUDA kernel for templates 

%In our implementation, each IIR filter is mapped onto one CUDA thread. 
%The challenge then, is to %to map SPIIR templates to CUDA blocks or CUDA warps to 
%maximize occupancy\footnote{Occupancy is defined as ratio of active warps number per multiprocessor to the maximum number of active warps possible. This provides a direct measure for the percentage of GPU utilization.} when the number of IIR filters between different templates can vary greatly.

%We map the whole SPIIR calculation (template array) onto the CUDA grid and one SPIIR filter onto one CUDA thread. 
%Each filter is mapped to exactly one CUDA thread.
%Considering the vast difference in size of SPIIR filters for different templates, we map the templates with more than 32 filters (warp size) into one block, and fit templates with less than 32 filters into one warp. 

%We design the Compute Unit function CU to calculate the minimum needed collaborated threads for a given SPIIR template $tp$ .
%\begin{equation}
%\label{equ:cu}
%CU(tp) = 
%\begin{cases}
%2^{t}, & 2^{t-1}< size(tp) \leq 2^{t} \land size(tp) \leq 32 \\
%32t,  & 32(t-1) < size(tp) \leq 32t \land size(tp) >32
%\end{cases}
%\end{equation}

The details of mapping of number of filters to CUDA threads is given here. We exploit the fact that CUDA threads are executed in groups of 32 (warp size), launching as much as possible our CUDA kernels with a \emph{number of threads = multiple of warp size}. This is considered to be optimal as it  helps avoid idled threads.
%simplifying the kernel code and avoiding branch execution paths. 
For a template with $N>32$ filters, we assign $M$ threads for this template, where $M$ is rounded to the next multiple of 32 after $N$.
For a template with  $N \le 32$ filters, we assigned $M$ threads where $M$ is rounded to the next power of 2 after $N$. 
Multiple templates may be executed in a warp if they are able to fit into the warp. 
For instance, a template with 513 filters will be assigned to 544 threads, while a template with 5 filters will be assigned to 8 CUDA threads.
Four small templates (with 5 filters) will be executed within a single warp. 
With this assignment, it can be guaranteed that the number of idle threads will not be more than 31 in the worst case. 

We use the \textit{Maxwell} architecture-based \textit{GeForce}GTX 980 GPU as our test machine, it has maximumly 2048 active threads, 
To maximize the active number of threads in each SM, 
the minimum number of threads for one block (the number of blocks per SM is 32) must be larger than $\frac{2048}{32}=64$. 
For our warp-based CUDA kernel, we chose to use 256 threads for each block as recommended in CUDA Programming Guide~\cite{nvidia2011nvidia}. 
For our block-based CUDA kernel, only one template is mapped to a single block. 
%These choices of thread numbers are able to help maximize the occupancy of Maxwell GPUs.
%Therefore the number of threads for each block is the closest multiple of warp size to cover the number of filters.
%$CU(tp)$ (from equation (\ref{equ:cu}) we know $CU(tp) \ge 64$) as the block size which means more threads will often be needed for larger template. 
%To improve the data access efficiency, we 
%also chose suitable amount of hardware resources for each thread. In CUDA, registers, used for storing local variables, are 
%pre-allocated registers to store local variables for each GPU threads. As the register size per SM is equal to 64 KB, to launch 2048 threads concurrently, the register memory assigned to each thread should not exceed $\frac{64}{2048}$KB$=32$ bytes, otherwise the threads will not be executed concurrently due to resource limitation. Similarly, available shared memory per SM is equal to 96 KB, so the maximum  shared memory assigned to each thread should not exceed  $\frac{96}{2048}$KB$=48$ bytes. 
%The Warp-based CUDA kernel will handle the case when the size of a template is not larger than 32. In this case, one or more templates can be mapped into one CUDA warp. The Block-based CUDA kernel will handle the case when the size of a template is larger than 32. In this case, one template will be mapped into one independent CUDA block.

\subsubsection{Read-only data cache}\label{sec:data_access} 

Data access is a critical aspect which can greatly affect the GPU performance. 
For unoptimized GPU programs, it may take longer to perform memory access than to do the core calculation. 
We analyzed the data access pattern of SPIIR filtering and applied an efficient data mapping method to improve the GPU performance. 
Three types of memory are investigated in our implementation. 

Register is the fastest accessible memory. 
It is local to individual threads and cannot be accessed by other threads. 
At the block level, there are shared memories accessible by all the threads of the same block (but not necessarily by the same warp). 
For the GTX 980 GPUs, one Maxwell streaming multiprocessor features 96 KB of dedicated shared memory.
Although shared memory features a broader memory access, it is much slower than the register, and usually requires synchronization within the block.
Global memory is located off the chip and its speed is the slowest in the GPU memory hierarchy. 
A new feature is introduced with the Kepler GK110 architecture (a predecessor of the Maxwell architecture)---the read-only data cache. 
It can be used to cache read-only global memory load which reduces the global memory access time. 
The GTX 980 has 4GB of global memory shared by all threads. 
To achieve high global memory throughput, we applied a coalesced memory access technique which can combine multiple memory accesses into as few as one cache transaction. 

\begin{figure}[htbp]
	\centering
	\includegraphics[width=12.0cm]{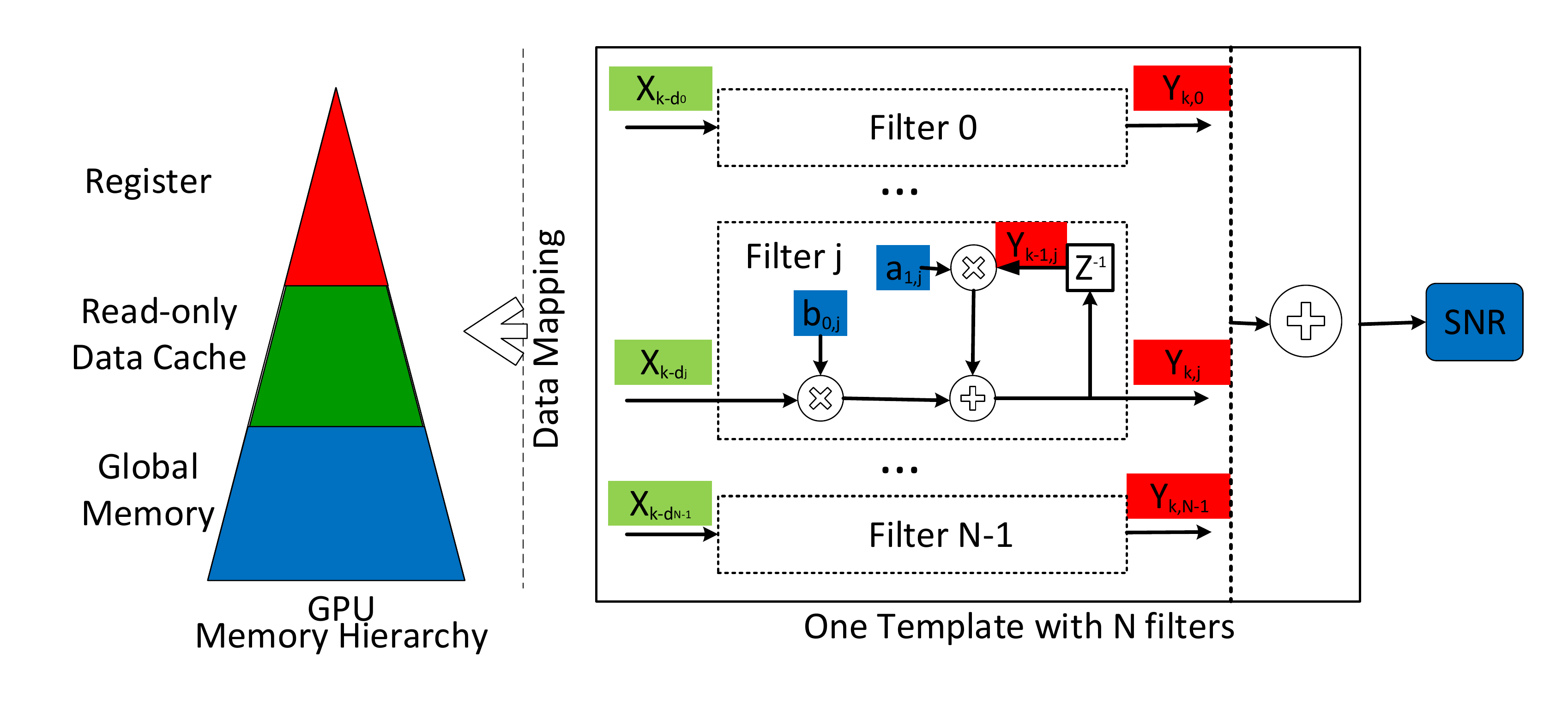}
	\caption{A schematic on how the data are mapped onto the GPU memory hierarchy to achieve low-latency data access for SPIIR filtering. The left part shows the color-coded GPU memory hierarchy. From top to bottom, the data access speed decreases but the data storage capacity increases. The right diagram illustrates the memory types created for variables for SPIIR filtering. $X_k$ is the input data to be filtered, $a, b$ are filter coefficients, $Y$ is intermediate output from each filter, $Z^{-1}$ represents the iterative process, and SNR is the filtering result.}
	\label{fig:memory}
\end{figure}

Fig.~\ref{fig:memory} is a schematic on how we organized and mapped the data of one SPIIR template to the GPU memory hierarchy in order to achieve low-latency data access. 
The output of the SPIIR method is given by Eq.~\ref{iireq0} and Eq.~\ref{iireq1}. 
Due to the iterative nature of the IIR filter, the left-hand-side of Eq.~\ref{iireq0}, $y_{k,j}$, will be reused by the next iteration. 
%Since there can be a huge number of iterations, reusing value of $y$ may drastically reduce the time of accessing memory.  
As there are a huge number of iterations, we therefore chose to store $y$ into a register to utilize the fastest memory access in GPU.
The input parameter $x_{k-d_{j}}$ cannot be reused by the same filter, but other filters of the same template may need to access it. 
The access pattern of $x_k$ has a good temporal and spatial locality so they are stored in the CUDA read-only data cache. 
Other input parameters, such as $a_{1,j}, b_{0,j}$, cannot be reused and so they are stored in the slower global memory, but we utilized the efficient coalesced memory access feature.
%To compute the final SNR ouput (Eq.~(\ref{iireq1})), we implemented a parallel sum reduction technique (see Sec.\ref{sec:synchronization}) that not only is efficient but also 
%to calculate the right-hand-side of Eq. (\ref{iireq1}). This method not only reduced the steps of sum operations from $(N-1)$ to $log_2N$ ($N$ is the number of filters in one template), but also used warp-shuffle to access the partial sum directly from different registers in the same warp. This hardware feature 
%greatly reduced the data access time.
Finally, the final SNR outputs are stored into global memory.

\subsubsection{Inner-warp and cross-warp summation optimization}\label{sec:synchronization}
%The synchronization among parallel threads often induces expensive overhead.
%However, in CUDA, the explicit synchronization operations may not be needed as the parallel threads can be organized into one warp. 
%The threads of one warp are synchronized automatically by GPU hardware with zero overhead (called implicit synchronization). 
%We applied this implicit synchronization feature of the CUDA warp to significantly reduce the cost of parallel threads synchronization.

%Atomic operations are often needed to access variables exclusively in a thread.
%For calculating summation with multiple threads, 
%simple implementation without atomic operations may yield wrong results. 
%For instance, in order to calculate $\sum_{n=1}^{32} N$,  will produce unpredictable result for each thread, as we i
%One therefore should avoid employing atomic operations as much as possible. 
%However, this understanding may be no longer true as for the latest GPU implementations. NVIDIA has significantly improved its atomic operation performance since Kepler GPUs, showing that global atomics can be as fast as typical global memory accesses \cite{nvidia2012gk110}. 

%The SPIIR method  allows all filters to execute their calculations independently, this is perfect to achieve parallel performance. However, for one SPIIR template with $N$ filters, we have to sum the output of $N$ filters to get the SNR. 
Obtaining the final SNR in  Eq.~\ref{iireq1} requires summation over all results of filters. This suggests a synchronization of outputting individual results when performing parallel computing. Previously we used the implicit synchronization feature of the CUDA warp to significantly reduce the cost of parallel threads synchronization and a multiple-thread parallel sum reduction method to reduce the cost of summation steps~\cite{liu2012gpu}.%a multiple-thread parallel sum reduction method to significantly reduce the summation steps.
%  in our warp-based CUDA kernel to calculate SNR and block-based CUDA kernel to calculate partial SNRs. 
%The parallel sum reduction method is first implemented by Liu et.~al.~in the SPIIR method~\cite{liu2012gpu} 
%and is now further optimized for Maxwell GPUs.
%The output of the parallel IIR filters of one template need to be synchronized in each iteration.
% this poses a big challenge for parallel threads collaboration because synchronization among parallel threads often leads to high latency. Fortunately,  every CUDA warp, e.g. 32 collaborated threads, are synchronized implicitly by the hardware with zero overhead thus no explicit synchronization operation is needed in warp-level. 
%This feature, if leveraged properly, may greatly improve application efficiency. 
%Further more, the new feature, warp-shuffle operation, allows the threads to access the registers of other threads in the same warp, which can also greatly improve the data access performance. 

Here the summation of all results within a warp is further improved by the warp-shuffle technique introduced from \textit{Kepler} michroarchitecture GPUs.
%as explained in Sec. \ref{subsec:memory} 
%to significantly improve the performance of parallel sum reduction. 
The warp-shuffle technique allows threads 
to read registers from other threads in the same warp. This is a great improvement over the previous higher-latency shared memory exchange within a warp.
%providing a new way for data sharing. is a feature introduced in that 

%For our warp-based CUDA kernel with templates $\le 32$ filters, w

%Method (a) of figure \ref{fig:reduction} is an example to show two SNRs can be generate directly by using the parallel sum reduction method. Two templates each with 16 filters will be mapped onto one warp and they require four sum steps to generate two SNRs. For each sum step, the implicit synchronization and warp-shuffle can greatly improve the performance of parallel sum reduction.

% For our Block-based-CUDA kernel (template size is larger than 32), we can also employ parallel sum reduction and warp-shuffle to get the partial SNRs of multiple warps. The next challenge is how to efficiently get SNR from the partial SNRs.

For our block-based CUDA kernel, where the template size is larger than 32, the output cannot be calculated without any cross-warp communication. %which inevitably introduces synchronization with overhead cost. 
%Therefore, the whole process is split into two phases. 
%First, each warp calculates one partial SNR. 
%Second, the partial SNRs are summed to calculate the final full SNR. 
%The first phase is exactly the same described above while
% naturally be conducted by leveraging the power of warp-shuffle operation. 
%The challenge is therefore to efficiently reduce the cross-warp communication overhead in the second phase.
%the second phase involves some cross-warp communication. 
We consider three different cross-warp summation methods to calculate the final SNR from partial SNRs.
The first one is the Direct Atomic summation (DA) method. It uses atomic operations of summation in global memory for all the partial SNRs.
This is the simplest and straightforward way to calculate SNRs from multiple partial SNRs. 
Atomic operations were considered expensive and ought 
to be avoided as much as possible before (without affecting the correctness of results). 
From \textit{Maxwell} GPUs, the atomic operations have been improved significantly. 

The second method is Shared memory Warp-shuffle (SW) method. It collects all the  partial SNRs of N times iterations into the shared memory of one warp (the batched computation model proposed in \cite{liu2012gpu}) and performs the warp-shuffle operation for the final SNR. Therefore, we need additional shared memory operations to calculate the final SNR based on the partial SNRs, one explicit synchronization operation to synchronize all the threads. 

The last method is Shared Memory Atomic Summation (SA) method. It collects partial SNRs into shared memory of one warp and performs atomic operation to compute the final SNR. This method also involves the same shared memory loading overhead and synchronization overhead as the SW method. Later we will show that the simplest atomic operations gives the best performance compared with the SW and SA methods. %We are able to simplify our implementation developed by Liu et.~al.~\cite{liu2012gpu} and improve our performance.

\subsection{Optimization of resampling and summation} \label{sec:resampling_addition}

\begin{figure}[htbp]
	\centering
%	\begin{subfigure}[htbp]{.4\textwidth}
		\begin{minipage}{.4\textwidth}
	\includegraphics[width=\textwidth]{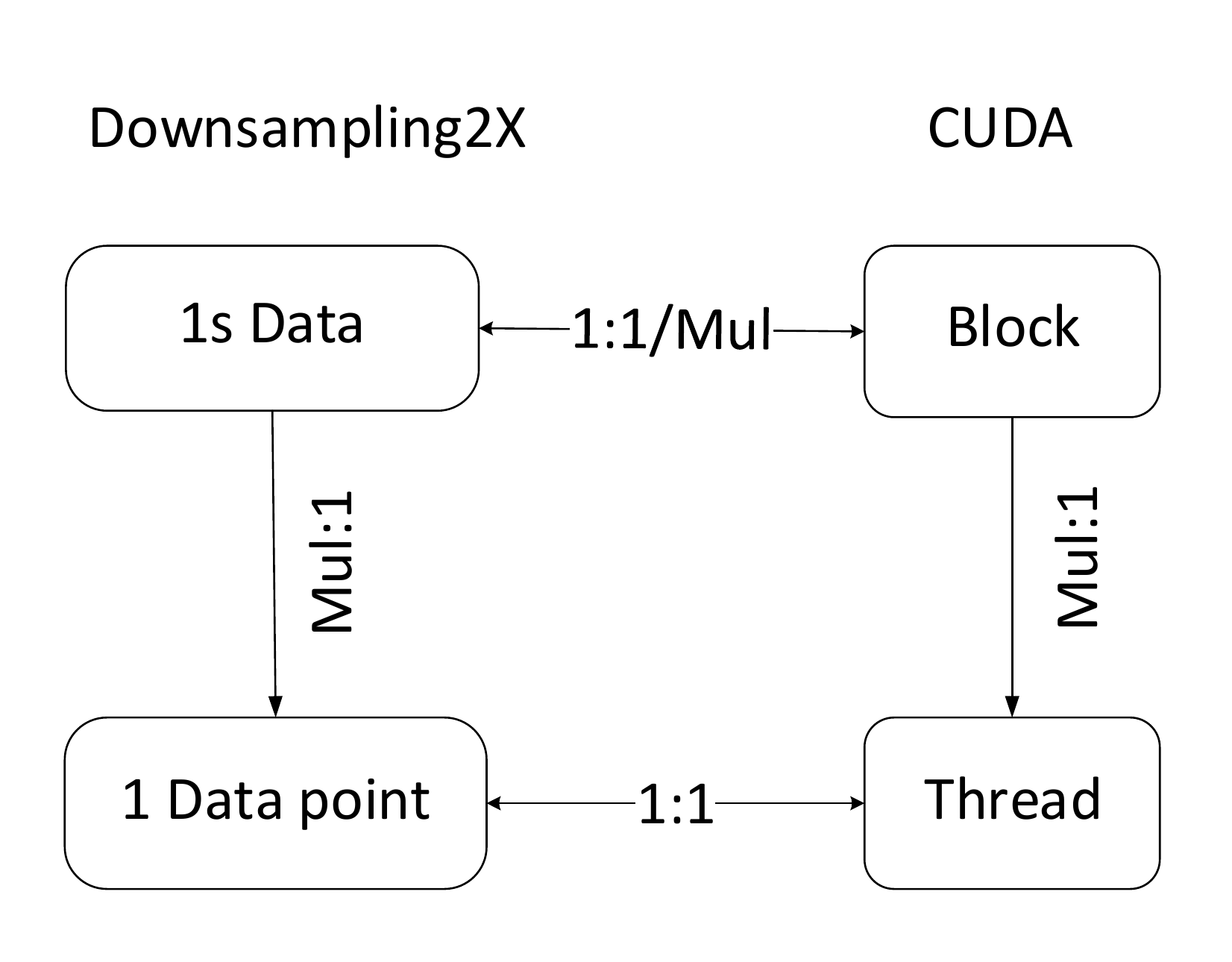}
	\caption{Schematic of the downsampling data hierarchy mapped onto the CUDA block-threads hierarchy. `Mul' means `Multiple', and '1s' means 'one second' in this figure.}
	\label{fig:down_mapping}
%	\end{subfigure}
\end{minipage}
\hspace{0.5in}
%	\begin{subfigure}[htbp]{.4\textwidth}
		\begin{minipage}{.4\textwidth}
	\includegraphics[width=\textwidth]{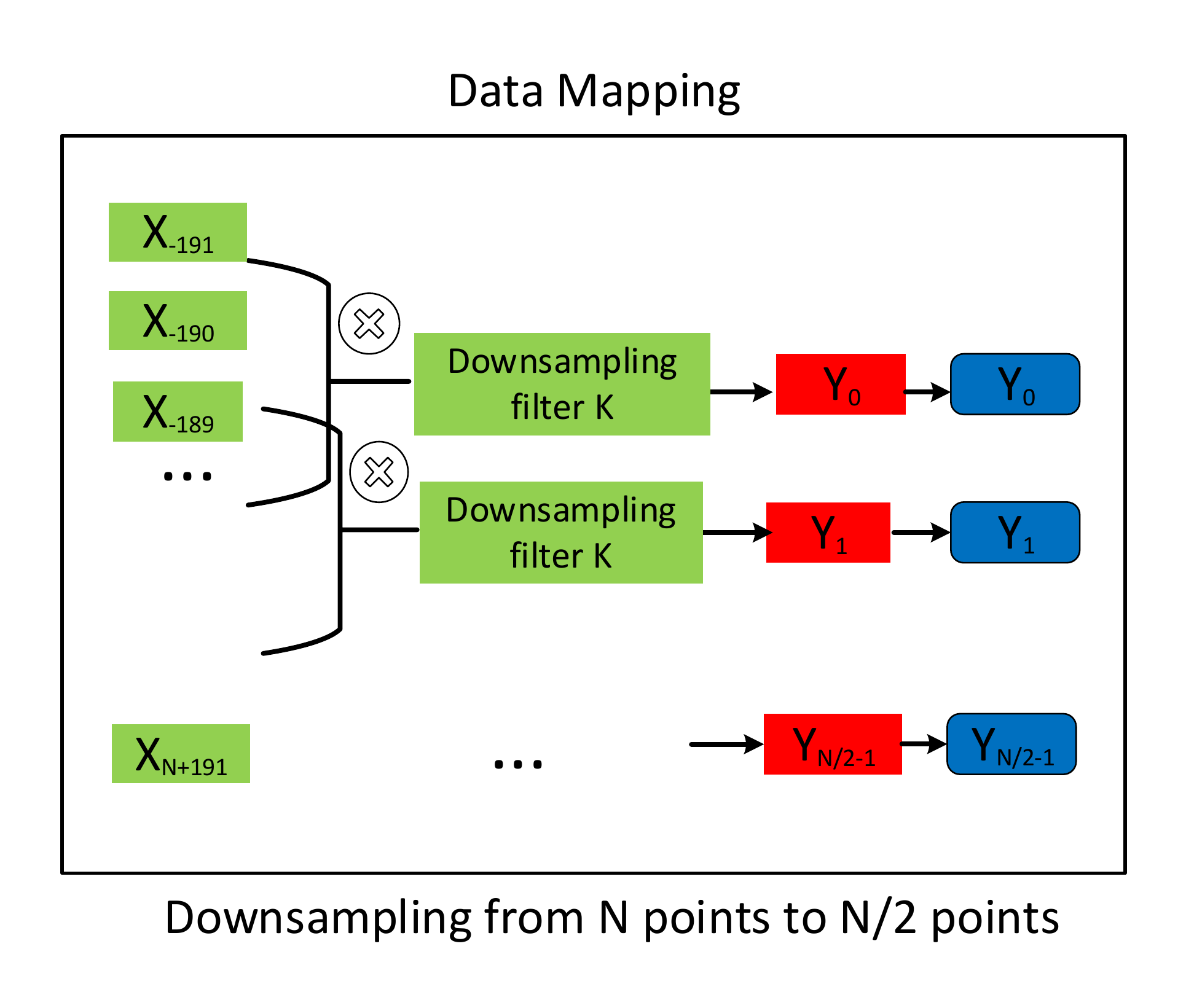}
	\caption{Schematic of the downsampling GPU kernel function and how data mapped onto the CUDA GPU memory hierarchy. Each color represent a type of GPU memory. Green: shared memory; red: register memory; blue: global memory.}
	\label{fig:down_memory}
%	\end{subfigure}
	\end{minipage}
\end{figure}

\begin{figure}[htbp]
	\centering
%	\begin{subfigure}[b]{.4\textwidth}
	\begin{minipage}{.4\textwidth}
	\includegraphics[width=\textwidth]{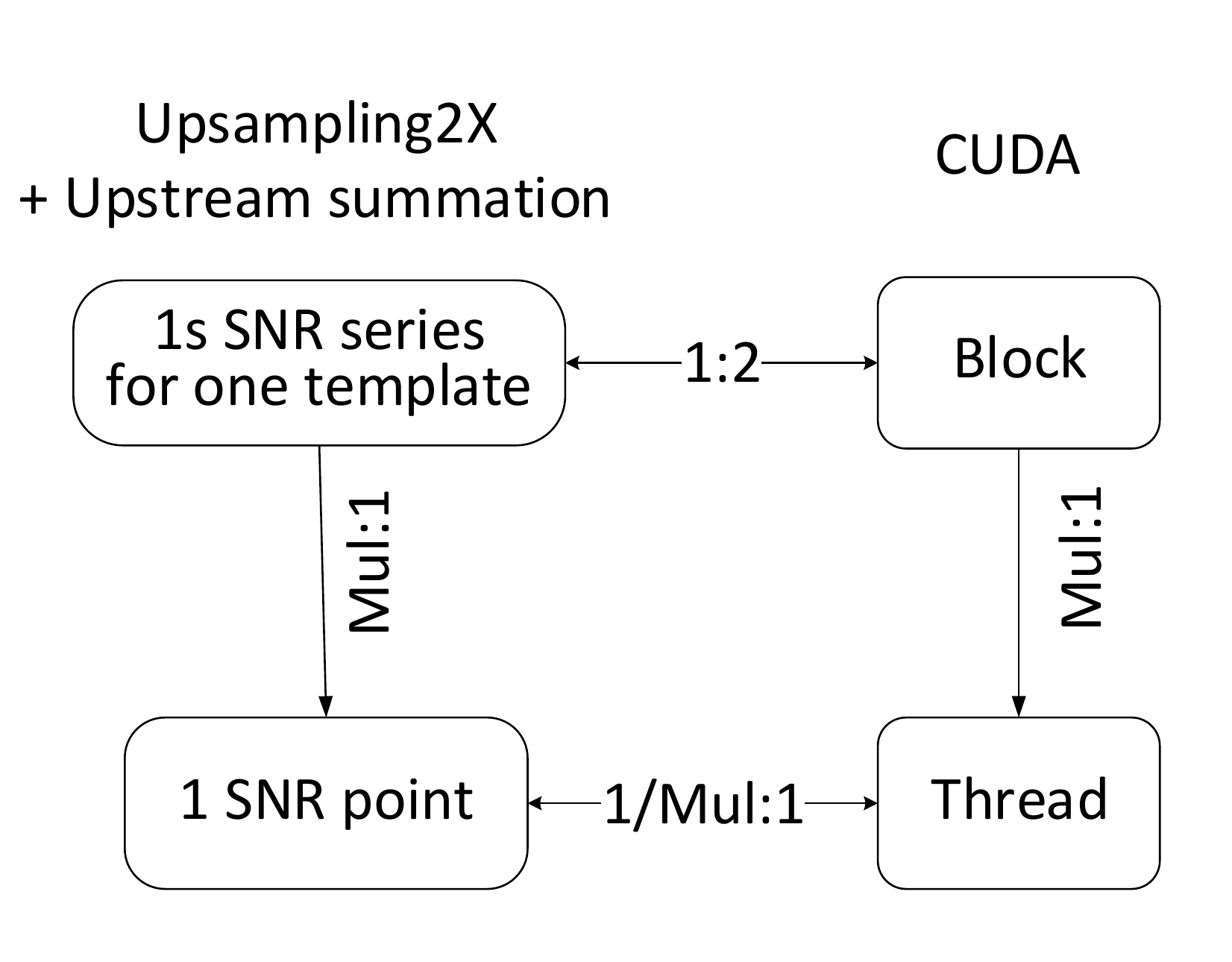}
	\caption{Schematic of the upsampling input data hierarchy mapped onto the CUDA block-threads hierarchy. `Mul' means `Multiple', and '1s' means 'one second' in this figure.}
	\label{fig:up_mapping}
	\end{minipage}
\hspace{0.5in}
%	\begin{subfigure}[htbp]{.4\textwidth}
	\begin{minipage}{.4\textwidth}
	\includegraphics[width=\textwidth]{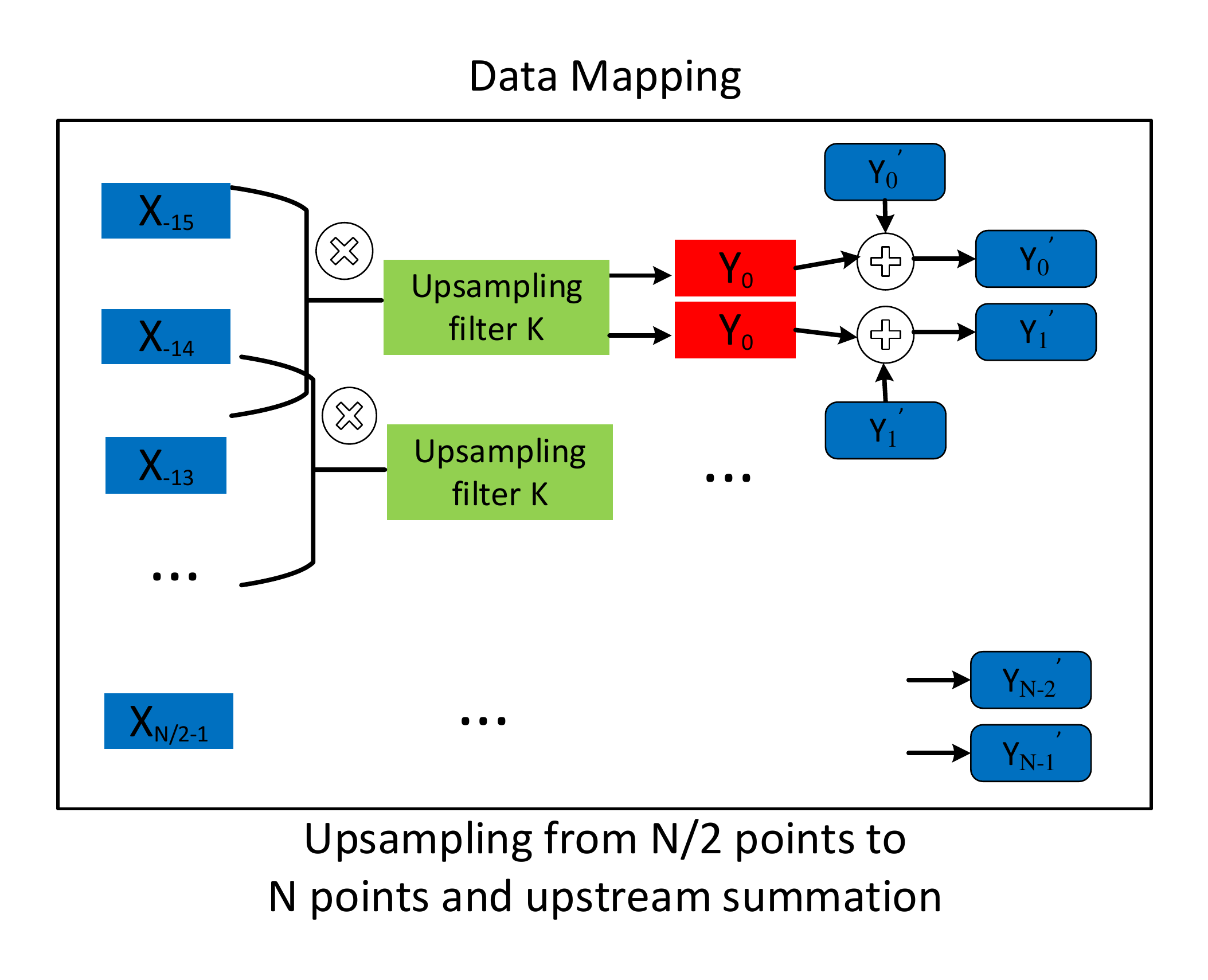}
	\caption{Schematic of the upsampling and combination GPU kernel function and how data mapped onto the CUDA GPU memory hierarchy. The left part shows the GPU memory hierarchy with explanation of the color code in Fig.~\ref{fig:down_memory}.}
	\label{fig:up_memory}
	\end{minipage}
\end{figure}

%In this section we accelerate the second computational part of the pipeline -- resampler and adder. Resampling is crucial for recovering a signal at arbitrary times given a set of discrete signal samples. This signal must comply the Nyquist-Shannon's sampling theorem in order to be correctly recovered.
We optimize the usage of CUDA threads and various types of memory as in Fig.~\ref{fig:down_mapping} and Fig.~\ref{fig:down_memory} for the downsampling process. As shown in Fig.~\ref{fig:down_mapping}, we map the production of one downsampled data point to one CUDA thread. The number of downsampled data points can vary over a power of 2 series ranging from 32 to 2048. For downsampling with output data points more than the number of GPU cores, the GPU cores will be fully occupied. It is inevitable for some sub-rate downsamplings, there will be idle GPU cores. As downsampling is the least computational process, that it is only performed once for all templates, in our multi-rate scheme, we do not further optimize the idle cores. The number of blocks allocated is designed as $\displaystyle \frac{N}{\min\{256,N\}}$ where $N$ is the number of downsampled points. For the three types of GPU memories, we map the input and downsampling Kaiser filter to the shared memory as they will be reused a number of times. Kaiser filtering is calculated iteratively and the intermediate filtering result is stored in the register memory for fastest data access. Finally, The global memory will store the downsampling output.

Similarly, Fig.~\ref{fig:up_mapping} and Fig.~\ref{fig:up_memory} show our CUDA design for the combined function of upsampling and upstream summation. As a one-second SNR series has a real and an imaginary component, we map each component of the SNR series to one CUDA block. Each thread is mapped to $\displaystyle \frac{2N}{\min \{256,N\}}$ upsampled SNR points where $N$ is the number total SNR points in a second. Different from the downsampling that only is perforce once, the upsampling and upstream summation needs to be performed for each template. The number of blocks will double the number of templates. The number of threads will be mostly likely more than the number of GPU cores, making a full occupancy of GPU cores. For the memory mapping, the upsampling Kaiser filter is mapped to the shared memory and the intermediate filtering output is mapped to the register memory, the same as the downsampling memory mapping. The input SNR series can not fit in the shared memory and they are stored in GPU global memory.

\section{Results of multi-rate SPIIR filtering}\label{sec:experiments}
\label{sec:er}
%The hardware we are using is a desktop shown in Tab.\ref{tab:testbed}. Note that the CPU implementation is compiled with compiler optimization \textit{O2}. Benchmarks show that the performance of \textit{O2} is almost always better than no optimization (\textit{O0})~\cite{compiler_O2}.   
In this section, we first show the GPU performance of SPIIR filtering and the improvement from each optimization step (Sec.~\ref{sec:result_spiir}). In the second part (Sec.~\ref{sec:result_multirate})., we show the GPU performance of several scenarios of multi-rate SPIIR filtering. This promises a lucrative performance improvement by GPU acceleration for our SPIIR pipeline. 

All GPU implementations are tested on a \textit{GeForce} GTX 980 (\textit{Maxwell} microarchitecture) GPU equipped desktop computer where Tab.~\ref{tab:testbed} shows its configuration. As the CPU counterparts are implemented in single CPU thread fashion, we use the elapsed time as our performance measurement to reflect the usage of CPU resource. 
%In the precision part, we study the precision issue that may arise on GPUs without ECC functionality. 
We run each experiment ten times and use the average time as the timing result. We set the number of templates to 4096 for performance testing purpose in this section. Note that the number of templates for a search can range from a few hundreds to hundreds of thousands. 
%$GPU\ prev$ refers to the SPIIR GPU implementation described in the work of Yuan. L. et al \cite{liu2012gpu}, while $GPU\ new$ stands for our Maxwell microarchitecture targeted, highly optimized SPIIR kernels.
\begin{table}[htbp]
\caption{Testbed configuration}
%\footnotesize
\centering
\begin{tabular}{ccc}
\hline
\multirow{3}{*}{Hardware} & CPU & Intel \emph{Core} i7-3770 3.40 GHz\\
& Host Memory & 8 GB DDR3\\
& GPU & NVIDIA \emph{GeForce} GTX 980\\
& GPU Memory & 4 GB DDR5\\
\cline{1-3}
\multirow{3}{*}{Software} & Operating System & Fedora 20 64-bit\\
& Host Compilation & gcc 4.8.3 -O2\\
& CUDA Compilation & nvcc 6.5\\
\cline{2-3}
\hline
\end{tabular}
\label{tab:testbed}
\end{table}

%A e comparison. The CPU SPIIR counterpart is  with -O2 optimization option.
%FIXME: table here

\subsection{Performance study of SPIIR filtering in single-rate}
\label{sec:result_spiir}

In this section , We first show the overall improved performance of our new GPU acceleration (noted as {\it New Kernel}) over the previous GPU acceleration \cite{liu2012gpu}  (noted as {\it Pre-Kernel}). We then show the breakdown of the improvement by exploiting new features of \textit{Maxwell} GPU.%we compare the performance of our new GPU acceleration (noted as {\it New Kernel}) with the previous GPU acceleration \cite{liu2012gpu}  (noted as {\it Pre-Kernel}) running on the same machine. We also show how our optimization methods can affect the performance.
%We first show our improved SPIIR kernel based on Maxwell microarchitecture (noted as {\it New Kernel}) with the previous SPIIR GPU kenel\cite{liu2012gpu}  (noted as {\it Pre Kernel}) and then show the breakdown of the improvement with different optimization strategies.  and compare the GPU performance with a CPU counterpart as well
\subsubsection{Varying number of filters}

{\it New Kernel}s and {\it Pre-Kernel}s are tested against the the filtering using a single-core CPU. Tab.~\ref{tab:spiir_speedup} shows the speedup ratio of {\it New Kernel}s in comparison to {\it Pre-Kernel}s with different number of filters for filtering. For number of filters $\geq 32$ where both kernels use the same thread configuration, the improvements by the {\it New Kernel}s over {\it Pre-Kernel}s increase as the number of filters increase, as shown by Tab.~\ref{tab:spiir_speedup}. This is mainly due to that the {\it New Kernel}s improve the data access and exchange speed, the effects of which are manifested with more number of filters used.  For the previous targeted optimization range (i.e. number of filters 128 to 256), the {\it New Kernel} have about 3-fold speed improvement over the {\it Pre-Kernel}. For the size of template (i.e. the number of filters) is relatively low (template size $<$ 32), our new thread configuration~---~warp-based kernel, improve the speed performance as the number of filters decrease, to a factor nearly 10-fold.  %The execution time for different number of filters is shown in Fig.~\ref{fig:exec-filters}.

\begin{table}[htbp]
\caption{Speedup ratios of different GPU kernels compared with the CPU counterpart.}
\begin{tabular}{c|c|cccccccc}
\hline
\multicolumn{2}{c|}{Template Size} & 4 & 8 & 16 & 32 & 64 & 128 & 256 & 512 \\
\hline
\multirow{2}{*}{Kernel} &$New Kernel$ & 63.10 & 70.62 & 61.22 & 55.98 & 109.76 & 124.41 & 125.58 & 123.97 \\
\cline{2-10}
% GPU v1 & 20.97 & 19.59 & 29.41 & 42.51 & 54.58 & 64.42 & 72.74 & 78.31 \\
&$Pre-Kernel$  & 6.13 & 10.79 & 19.61 & 37.50 & 42.64 & 48.72 & 39.51 & 29.78 \\
\hline
\end{tabular}
\label{tab:spiir_speedup}
\centering
\end{table}

\subsubsection{Performance Improvement by Employing Warp-shuffle and Read-Only Data Cache}  

%To implement the parallel sum reduction in one warp, one thread will need the data from another thread's register. In the early Fermi microarchitecture, we must store the result to shared memory first, and then another thread can load it back from shared memory to calculate the summation. However, in the latest Maxwell microarchitecture, which can support warp-shuffle, one thread can directly get the data from another thread's register (in the same warp). So employing warp-shuffle can achieve better data access performance. 

\begin{figure}
\centering
\begin{minipage}{.49\textwidth}
  \centering
  \includegraphics[width=.9\textwidth]{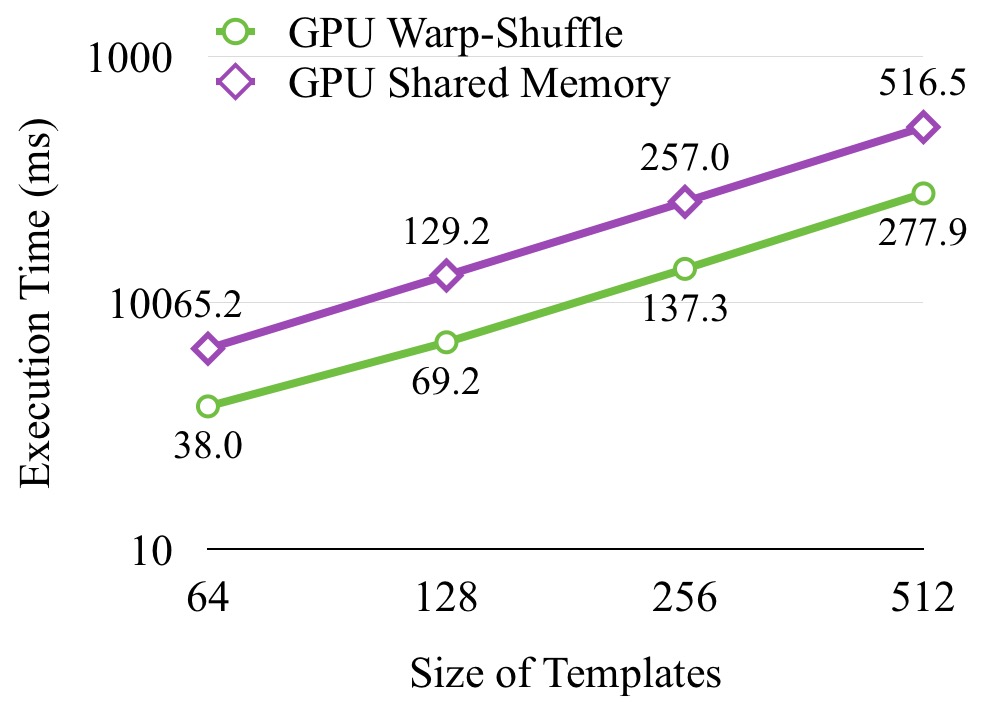}
  \captionof{figure}{Performance comparison in performing parallel summation reduction between shared memory and warp-shuffle using 4096 templates}
  \label{fig:shared-warpshuffle}
\end{minipage}
\hfill
\begin{minipage}{.49\textwidth}
  \centering
  \includegraphics[width=.9\textwidth]{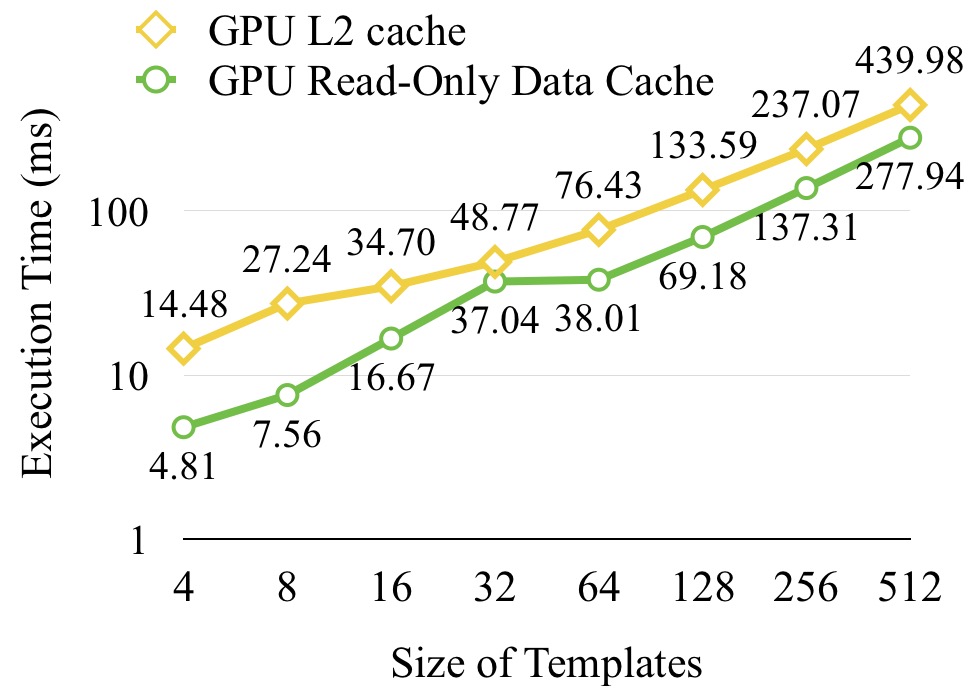}
  \captionof{figure}{Memory access performance comparison between caching parameter $x$ in L2 cache and read-only data cache using 4096 templates}
  \label{fig:L2-readonly}
\end{minipage}
\end{figure}

Fig.~\ref{fig:shared-warpshuffle} shows the performance of two different implementations with and without warp-shuffle. The first implementation uses warp-shuffle to access data within a warp to calculate the summation. The second implementation uses shared memory to access data and calculate summation. It shows that warp-shuffle summation performs significantly faster than shared memory summation.
% which justifies our optimization choice.

Fig.~\ref{fig:L2-readonly} shows the effect of using the read-only data cache. The \textit{Maxwell} GPU loads its global memory data to L2 cache rather than to L1 cache. In Fig.~\ref{fig:L2-readonly}, GPU L2 cache illustrates the kernel performance when storing parameter $x$ of Eq.~\ref{iireq0} in the L2 cache, which is the default cache for global memory.  Because variables such as $x$ of Eq.~\ref{iireq0} has a favorable temporal and spatial locality, using the read-only data cache can be very efficient as shown in the figure.

\subsubsection{Performance Comparison Among Three Cross-warp Summation Methods} 

\begin{figure}
\centering
\begin{minipage}{.49\textwidth}
  \centering
  \includegraphics[width=.9\textwidth]{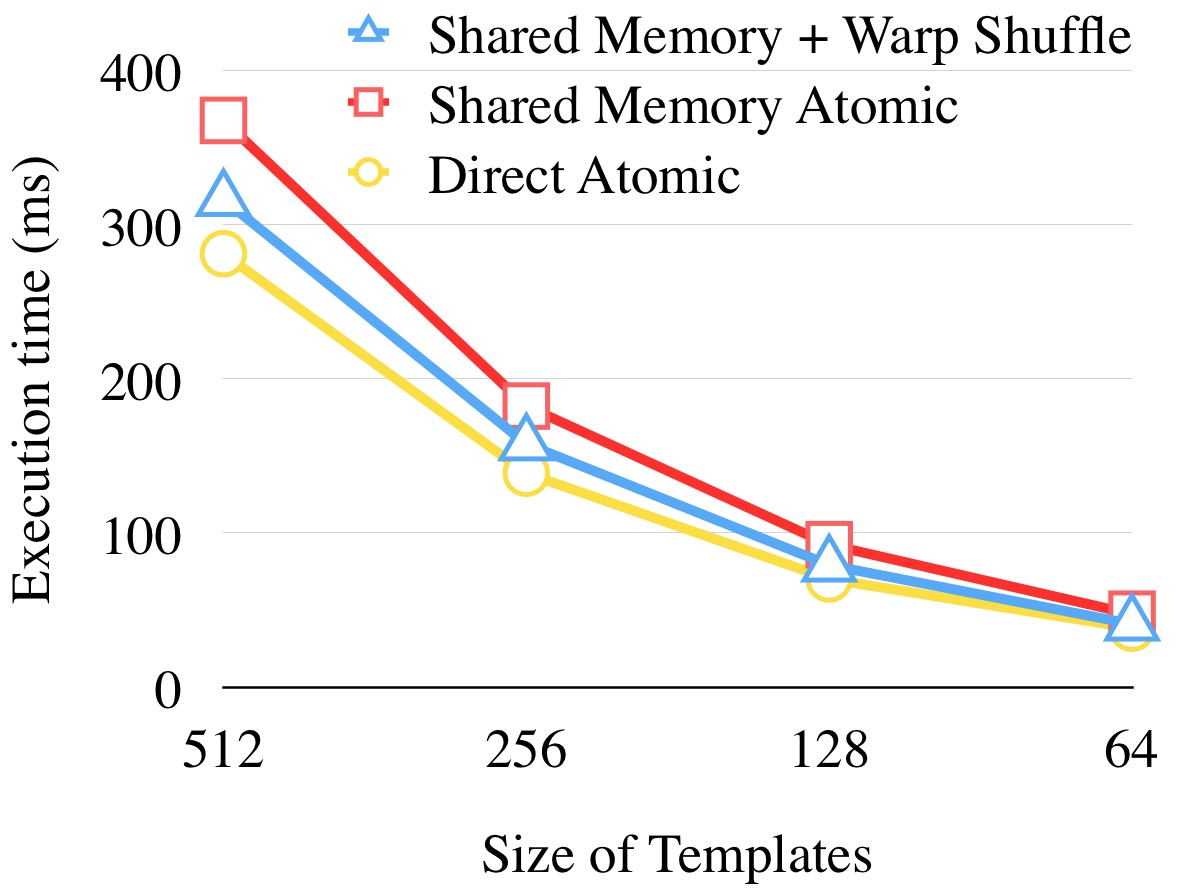}
  \captionof{figure}{Execution time of different cross warp summation methods with 4096 templates in different execution configurations.}
  \label{fig:atomic-shared}
\end{minipage}
\end{figure}

Fig.~\ref{fig:atomic-shared} illustrates the performance of the block-based SPIIR kernels using three different cross-warp summation methods:
DA (directly using atomic operation in global memory),SW (using warp-shuffle and shared memory) and SA (using atomic operation in shared memory) methods.
%Direct Atomic Summation (DA), Shared Memory Warp-Shuffle (SW), and Shared Memory Atomic Summation (SA).
% (methods (b), (c), (d) in Fig. \ref{fig:reduction}). 
To ensure objective comparison, we attempt to optimize each implementation as much as possible. GPU bank conflicts are perfectly avoided in the SW and SA methods where shared memory usage is involved. As one would expect that shared memory access is much faster than global memory access, the SW and SA methods could be faster than DA method. Surprisingly, the DA method surpasses the other two methods as shown in Fig.~\ref{fig:atomic-shared}.

\begin{figure}
	\centering
	\begin{minipage}{.49\textwidth}
		\centering
		\includegraphics[width=.9\textwidth]{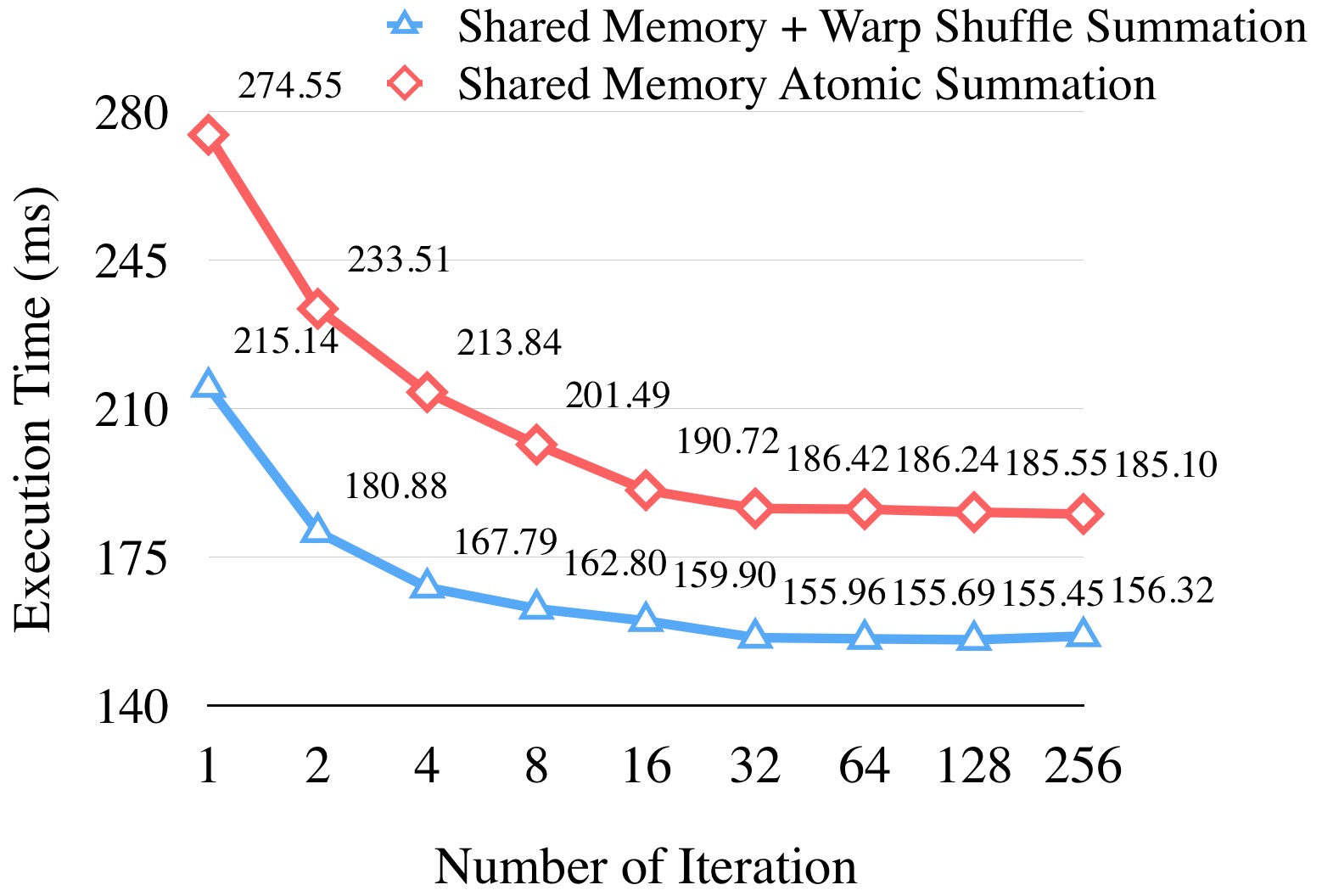}
		\captionof{figure}{The optimization effect when we change the iteration number. Template size of 256 is used (with 4096 templates).}
		\label{fig:batch}
	\end{minipage}
	\hfill
	\begin{minipage}{.49\textwidth}
		\centering
		\includegraphics[width=.9\textwidth]{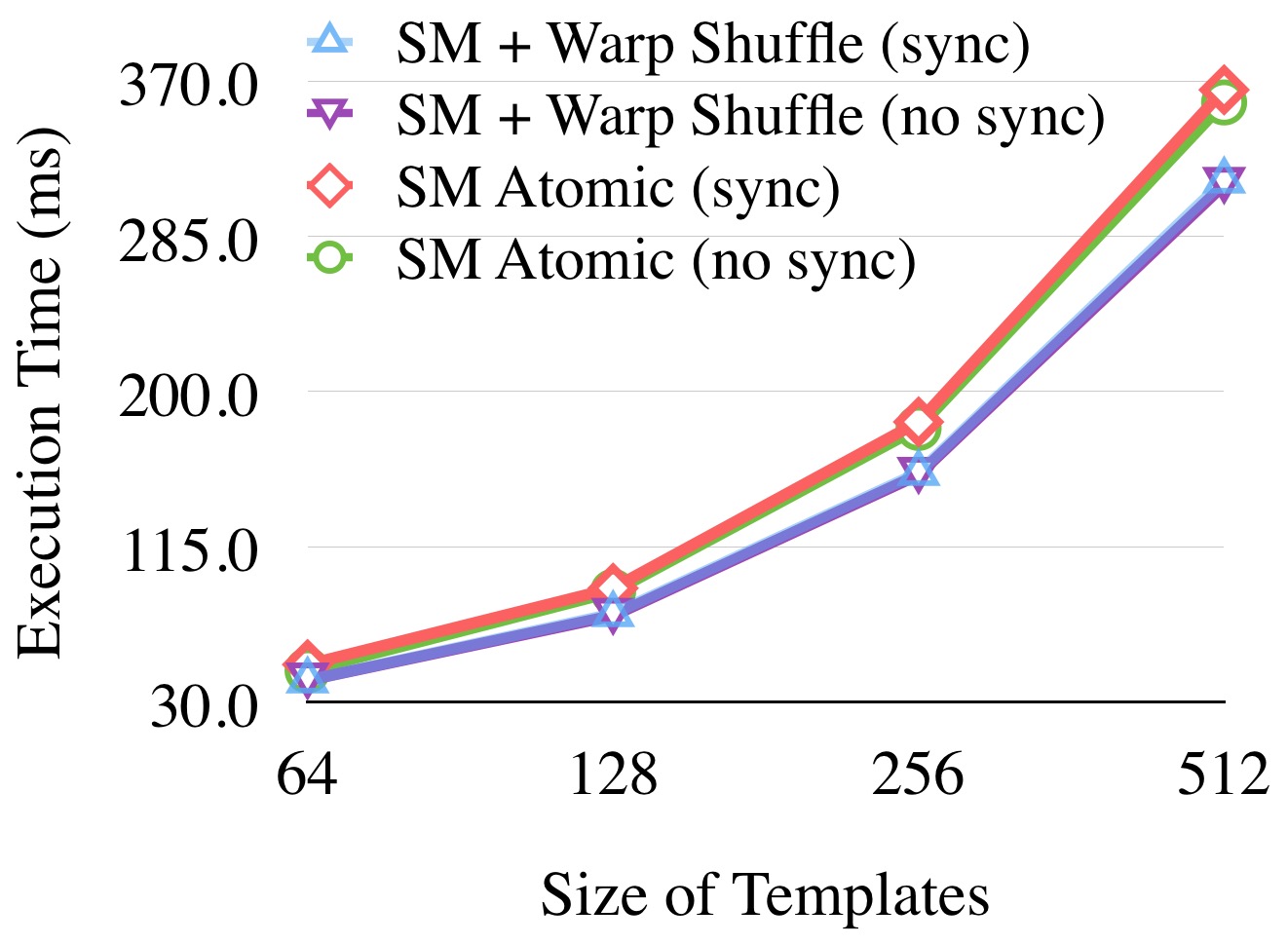}
		\captionof{figure}{The overhead of explicit synchronization can be ignored when we use large iteration number. Iteration number 256 is used here (with 4096 templates)}
		\label{fig:sync-influ}
	\end{minipage}
\end{figure}

We design two additional experiments to figure out the reason for the inferior performance of the shared memory methods (SW and SA), whether it is because of the cost of  shared memory accesses or the explicit synchronization. To reduce the total synchronizaiton cost, both SW and SA methods take advantage of shared memory to execute many iterations before one explicit synchronization operation. Fig. \ref{fig:batch} clearly illustrates that the cost of synchronization can be significantly reduced or amortized into many iterations. The larger the iteration number, the better the performance. However, when the iteration number reaches a certain point, the performance benefit becomes almost unchanged as the iteration number continue growing. Fig.\ref{fig:sync-influ} shows that the reason is that the synchronization overhead has been completely hidden by calculation in such circumstances. In Fig.\ref{fig:sync-influ} we disabled the synchronization operation in these shared memory GPU kernels to observe the performance influence of explicit synchronization operations in sufficiently large iteration number. Though the computation result may not be correct, the comparison itself does show the impact of explicit synchronization operations. As can be seen, the performance between synchronization and no synchronization GPU kernels are almost unidentifiable, illustrating that we can completely remove the overhead of explicit synchronization by improving the iteration number. Therefore, the additional shared memory operations introduced in SW and SA methods lead to their lower performance compared with DA method.  %Combining the result from Fig. \ref{fig:atomic-shared}, we observe that warp shuffle operations outperforms shared memory regarding to latency, though they consist of SRAMs.  

% (we will show that the cost of explicit synchronization is negligible because we remove this cost by using large numbers of SNRs). 
%so the simplest and low cost direct atomic summation method DA wins. 
%Fig. \ref{fig:atomic} shows that atomic operations in our pipeline do not affect the performance significantly. This means the latency of atomic operations can be concealed as long as the occupancy is high enough, which in turn can be achieved by our pipeline optimization. 
%For method shared memory atomic, although this method is able to reduce explicit synchronization time, it is very unlikely that the atomic operation latency in shared memory can be perfectly hidden since shared memory has much higher bandwidth than global memory.

\begin{figure}
  \centering
\begin{minipage}{.49\textwidth}
  \includegraphics[width=.9\textwidth]{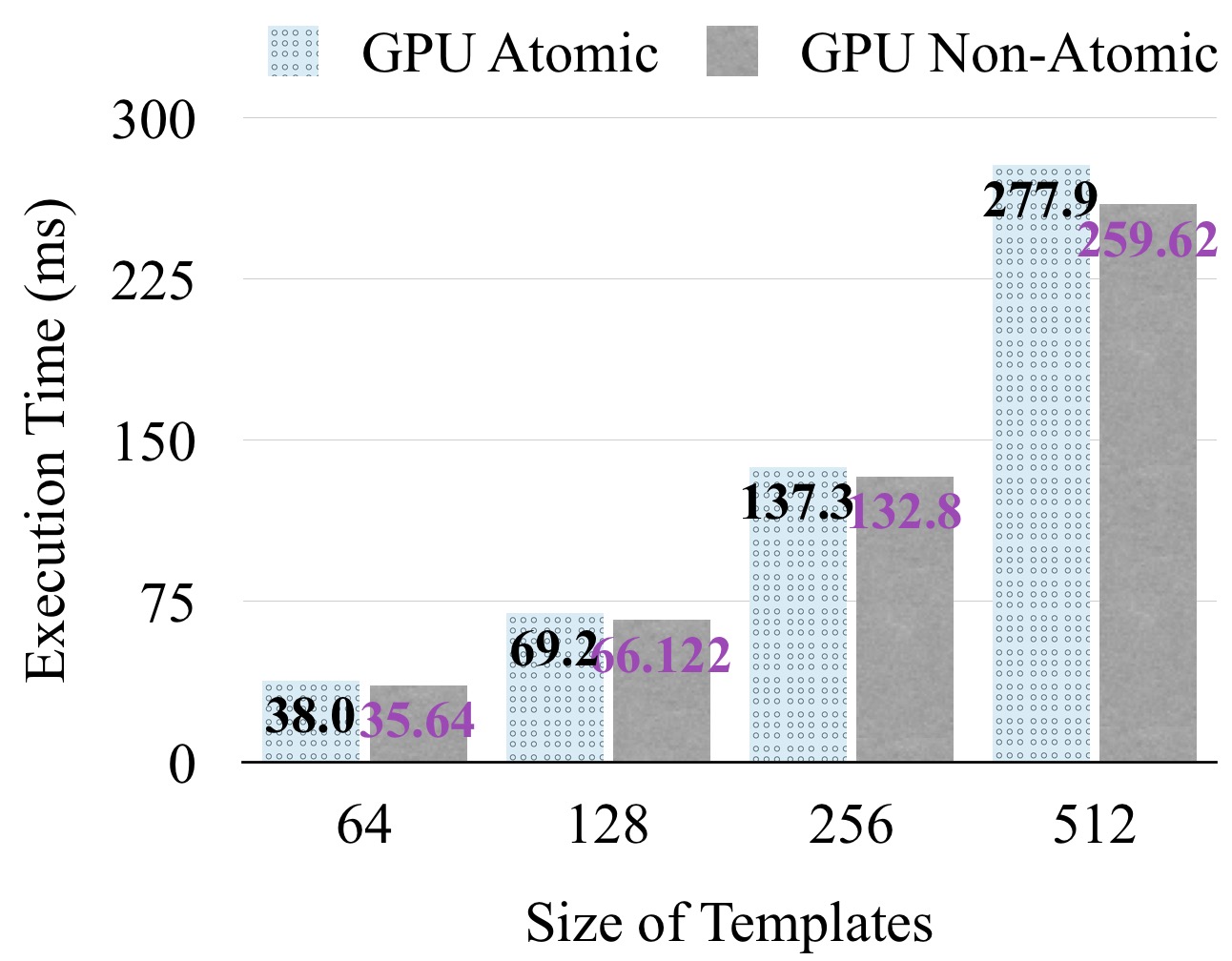}
  \captionof{figure}{Execution time of atomic and non-atomic kernels with 4096 templates in different execution configurations.}
  \label{fig:atomic}
\end{minipage}
\end{figure}

The impact of atomic operations is also tested for the DA method. Atomic operations are typically considered costly and should be avoided whenever possible, however Fig.\ref{fig:atomic} shows the cost of atomic operation can be almostly ignored because the execution time of two kernels with and without atomic operations is so close. All the experiments explain why DA method is better than SW and SA methods. 
%We know that the kernels without atomic operations can not guarantee the correct results but we can really compare the performance. As can be seen, the atomic version is only marginally slower than the non-atomic version. %This is due to the very low atomic operation latency achieve by the new GPU microarchitecture, and the result of latency hiding benefited from very high occupancy (100\%).

\subsection{Performance of SPIIR filtering in multi-rate}
\label{sec:result_multirate}
 
The results shown in the last section is SPIIR filtering in single-rate. Here we show the performance of the GPU-accelerated multi-rate implementation of SPIIR filtering, which includes GPU accelerated rate alterations. We test several multi-rate scenarios denoted as "number of rates"  in Tab.~\ref{tab:multirate_speedup}. ``Number of rates" as 1 in the table denote the single rate at full rate (4096 Hz). The number of filters at full rate for the 4096 templates is set as 1024. The sub-rates are formed according to the design shown in Sec.~\ref{sec:multi_design}. The initial filters are equally divided to each sub-rate for testing purpose. Tab.~\ref{tab:multirate_speedup} shows elapsed times for the CPU and the GPU-accelerated multi-rate filtering in different scenarios. The efficiency of our multi-rate design shown here is consistent with our estimation in Sec.~\ref{sec:multi_design} that the 8-rate scenario reduces the measured time to around one quarter. Our GPU acceleration of multi-rate scheme is dominated by the GPU acceleration of SPIIR filtering in single-rate. Therefore it is very efficient that we are able to obtain more than 100-fold speedup for any scenario.

\newsavebox{\boxmultiratespeedup}
\begin{lrbox}{\boxmultiratespeedup}

%\begin{tabular}{p{5cm}|p{1.5cm}|p{1.5cm}|p{1.5cm}|p{1.5cm}|p{1.5cm}}
	\begin{tabular}{c|c|c|c|c|c}
\hline
Number of rates		 & 1 & 2 & 3 & 4 & 8 \\
%\hline
%CPU SPIIR kernel	 & 34815.2319 & 26187.2173 & 20517.5418 &  16342.1789 & 8453.0116  \\
\hline
CPU multi-rate scheme & 34815s  & 26379s  & 20817s  & 16680s  & 8841s \\
SPIIR filtering & 34815s  & 26187s  & 20517s & 16342s & 8453s \\
\hline
GPU multi-rate scheme	& 287s & 220s & 199s & 143s &  85s \\
\hline
Speedup & 121x & 120x & 104x & 116x & 104x \\
\hline
\end{tabular}
\end{lrbox}

\begin{table}[htbp]
\centering
	\caption{Elapsed times of CPU and GPU-accelerated multi-rate filtering scheme in different scenarios. }
	\label{tab:multirate_speedup}
			\scalebox{0.8}{\usebox{\boxmultiratespeedup}}
\end{table}

%[ 26187.2173  26379.5079  20517.5418  20817.9944  16342.1789  16680.5009
%   8453.0116   8841.8207]

%>>> 34815.2/287.4
%121.13848295059151
%>>> 26379.5/220
%119.90681818181818
%>>> 20818.0/199.6
%104.29859719438878
%>>> 16680.5/143.6
%116.15947075208913
%>>> 8841.8/85.1
%103.89894242068155

\section{Results of the GPU-accelerated low-latency SPIIR pipeline} \label{sec:pipeline}
%

\iffalse
\begin{figure}[htbp]
\centering
\includegraphics[width=10.0cm]{pie.pdf}
\caption{Computation profile of the low-latency SPIIR coincident search pipeline \textit{gstlal\_iir\_inspiral} performing on the recolored LIGO S5 data. The data overall is a representation of the clean detector data. There are reasonable amount single events for coincident analysis which mainly accounts for "Others". }
\label{fig:pie}
\end{figure}
\fi
%
\begin{figure}[htbp]
\centering
\includegraphics[width=15.0cm]{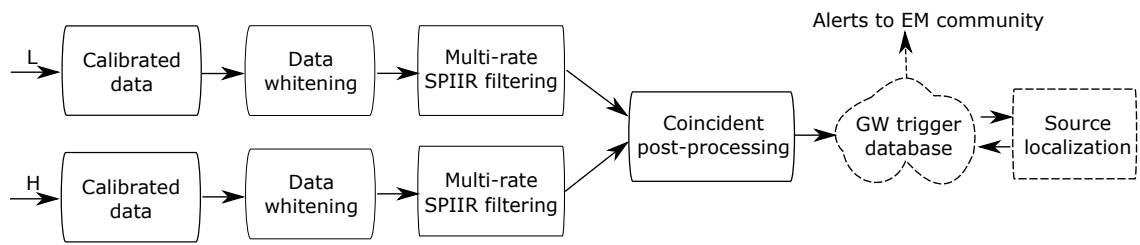}
\caption{Schematic flowchart of the low-latency SPIIR pipeline \textit{gstlal\_iir\_inspiral} with data input from two detectors. Solid blocks depict main components of this pipeline. Dashed blocks depict external processes.}
\label{fig:pipeline}
\end{figure}

The low-latency SPIIR detection pipeline we are considering here is publicly available through distribution of \textit{gstlal} software library~\footnote{\textit{gstlal} library: https://wiki.ligo.org/DASWG/GstLAL}. \textit{gstlal} provides a variety of components from LIGO Algorithm Library (LAL)~\footnote{\textit{lal} library: https://www.lsc-group.phys.uwm.edu/daswg/projects/lalsuite.html} for LIGO data processing and uses the \textit{gstreamer} framework to control streaming data. We use the existing \textit{gstlal} components for reading calibrated data, data whitening, performing coincident post-processing for our pipeline. We use our own multi-rate SPIIR filtering for template filtering. The pipeline has the code name \textit{gstlal\_iir\_inspiral} in \textit{gstlal}. A schematic flowchart of the pipeline is shown by Fig.~\ref{fig:pipeline}.

%It will search through individual single events for coincident events as GW candidates. The significance of a candidate is justified by the uncoincident events. The candidates can be submitted to the GW database where follow-up actions can be taken. 

The \textit{gstreamer} framework inherently employs multi-threading technique to take advantage of the multiple cores of a CPU. Therefore, the CPU implementation of the SPIIR pipeline is by default multi-threaded. We propose a different criterion, the CPU time used by all CPU cores, rather than the elapsed time for the single-threaded CPU implementation, to measure the performance of the GPU accelerated pipeline versus the original CPU pipeline.

\begin{table}[htbp]
	\caption{Computation profiling of the CPU low-latency SPIIR pipeline \textit{gstlal\_iir\_inspiral}.}
		\centering
		\begin{tabular}{c|c|c|c}
			\hline
			\multicolumn{3}{c|}{Multi-rate SPIIR filtering} & \multirow{2}{*}{Other} \\
			\cline{1-3}
			 SPIIR filtering & Resampling & Upstream summation & \\
			\hline
			 93\% & 3.1\% & 0.9\% & 3\% \\
			\hline
		\end{tabular}
\label{tab:pipeline_profile}
\end{table}

We present the used CPU time of the main components of the CPU pipeline in percentages, shown by Tab.~\ref{tab:pipeline_profile}. This computation profiling is performed using the platform given by Tab.~\ref{tab:testbed}. The data we process is a short segment of recolored LIGO 5th Science run data, which represents a set of clean data and produces a reasonable number of single events for coincidence analysis. The SPIIR filtering dominates the computation, taking over $93\%$ of the CPU time while the resampling and summation come next at $4\%$.

We now can estimate the expected CPU resource reduction of the pipeline, if part of the whole the multi-rate filtering component is executed on GPU instead of CPU. The expected resource reduction $\kappa_{\text{app}}$ is determined by the resource reduction brought by accelerated module $\kappa_{\text{mod}}$ and the computational cost of this module $P_{\text{mod}}$ in the pipeline. It is given by:
\begin{equation}
\kappa_{\text{app}} = \frac{1}{1-P_{\text{mod}}+P_{\text{mod}}/\kappa_{\text{mod}}},
\label{eq:pipeline_speedup}.
\end{equation}
If we only apply GPU acceleration only on the module of SPIIR filtering in our pipeline and we choose a somewhat $100$-fold from Tab.~\ref{tab:spiir_speedup} for this component. The total resource reduction ratio of our pipeline will be not more than $13$-fold. If on top of that, we apply GPU acceleration on resampling and upstream summation components, and we choose a somewhat $100$-fold from Tab.~\ref{tab:multirate_speedup} for this multi-rate SPIIR filtering. The total resource reduction of our pipeline will be about $25$-fold. 
%where $\kappa_{\text{app}}$ is the resource reduction ratio of the GPU-accelerated application, $\kappa_{\text{mod}}$ is the acceleration ratio of the accelerated module. $P_{\text{mod}}$ and $P_{\text{rest}}$ represent the computation percentage of the accelerated module and the rest modules of the pipeline.

We measure the CPU resource reduction efficiency and also the elapsed time gain using GPU acceleration on multi-rate SPIIR filtering. The setup of the performance test is explained in detail here. We simulate two sets of data using Advanced LIGO noise for the twin LIGO detectors, respectively. The duration of each data set is 1000 seconds. These data sets are injected coherently into a simulated binary neutron star coalescence GW signal. To prepare the templates used by our pipeline for the search, we generate two template banks for each detector, which covers the parameters of the injection. Each bank consists of 1024 geometric templates. We generate SPIIR filters for each bank and divided the filters into four groups corresponding to four designated sub-rates. We insert filters with zero coefficients into each group so that the number of filters of each group in a bank will be a power of 2. While it is not necessary to do the filter insertion for the search, the insertion here is for performance testing purpose. The number of SPIIR filters of each group in each bank is shown in table~\ref{tab:bank_config}.

\newsavebox{\boxbankconfig}
\begin{lrbox}{\boxbankconfig}
	\begin{tabular}{c|c|c|c|c}
		\hline
		Sampling rate (Hz) & 4096 & 2048 & 1024 & 512 \\
		\hline
		Frequency band (Hz) & $<2048$ & $<1024$ & $<512$ & $<256$ \\
		\hline
		Number of filters & 64 & 64 & 128 & 256 \\
		\hline
	\end{tabular}
\end{lrbox}
	
\begin{table*}[thbp]
	\begin{center}
	\caption{Number of filters in each frequency band of a bank and the data sampling rates required by Nyquist theorem.}
	\label{tab:bank_config}
	\scalebox{0.8}{\usebox{\boxbankconfig}}
				
	\end{center}
\end{table*}

\newsavebox{\boxpipelinespeed}
\begin{lrbox}{\boxpipelinespeed}
%	\begin{tabular}{p{2cm}|p{3cm}|p{3cm}|p{3cm}|p{2cm}}
	\begin{tabular}{c|c|c|c|c}
	\hline
	Pipeline & Used CPU time & CPU resource reduction ratio & Elapsed time & Speed-up \\
	\hline
	CPU pipeline & 31610s & 1x & 4560s & 1x\\
	%hline
	%GPU pipeline version 1 & {3110s} & {10x} & 640s & 7x\\
	\hline
	GPU pipeline & {1520s} & {21x} & 430s & 11x\\
	\hline
	\end{tabular}
%	\caption{Performance of the low-latency SPIIR pipeline merely using CPU power (CPU pipeline) and with GPU acceleration of multi-rate SPIIR filtering (GPU pipeline).}
%\label{tab:pipeline_speed}
%\centering
\end{lrbox}

\begin{table*}[htbp]
	\begin{center}
	\caption{Performance of the low-latency SPIIR pipeline merely using CPU power (CPU pipeline) and with GPU acceleration of multi-rate SPIIR filtering (GPU pipeline).}
		\label{tab:pipeline_speed}
		\scalebox{0.8}{\usebox{\boxpipelinespeed}}
	\end{center}
\end{table*}

%We measure the usages of CPU time and the actual elapsed time for the CPU pipeline, the pipeline that only accelerates the SPIIR filtering (GPU pipeline version 1), the pipeline that accelerates the multi-rate SPIIR filtering (GPU pipeline version 2) shown by 

Tab.~\ref{tab:pipeline_speed} shows the CPU resource reduction and elapsed time gain by the GPU-accelerated SPIIR pipeline. It achieves $ 21$-fold improvement on CPU resource reduction. This is close to our expectation shown earlier. Besides, we have significantly reduce the running time of the pipeline by 11-fold. The difference of the SNRs between the GPU pipeline and the CPU pipeline on the injection event is within $0.00006\%$ for the 10 test runs, and the injection is successfully recovered by the GPU pipeline.%The speed-ups are measured as $7\times$ and $11\times$.

\section{Conclusions and Future Work}
\label{sec:con}
Low-latency and real-time detections of gravitational-wave (GW) signals are gaining priority for their potential to enable prompt electromagnetic follow-up observations. The low-latency SPIIR detection pipeline is a CBC detection pipeline with that latencies of tens of seconds. %The SPIIR filtering is uses a very good parallelism that allows such real-time requirements to become possible.

%The GPU work of SPIIR is first introduced in 2012. As time goes, the once efficient GPU SPIIR implementation is not as efficient in newer hardware and new execution scenarios, to address this issue, 
In this paper, we develop the GPU acceleration for the main computational component of the SPIIR pipeline~---~multi-rate SPIIR filtering. We first improve the GPU optimization of the filtering part, by employing a new kind of GPU thread configuration and exploiting new memory features of \textit{Maxwell} GPUs. This provides a notable $1.5$ to 10 fold speedup over our previous acceleration on \textit{Fermi} GPUs. We implement GPU acceleration of resampling and upstream summation parts for the multi-rate filtering procedure. Our tests show the multi-rate filtering has been accelerated over 100 fold given different filtering scenarios. This leads to a 21-fold CPU resource reduction for the entire pipeline and a 11-fold reduction on elapsed time.

%Through these combined efforts, the SPIIR method can use the complete and full power of NVIDIA's latest Maxwell GPU. 

%We develop GPU acceleration for a multi-rate SPIIR filtering scheme. We incorporte the GPU acceleration into the low-latency SPIIR pipeline and achieve a near-optimal resource reduction rate. This GPU-accelerated pipeline has been used for LIGO engineering runs and science runs, including the first science run.

%We show that the latest GPU can be approximately 100\% utilized and a maximum speedup of 125 can be achieve. We can process 23000 templates in real time. This new progress makes real-time GW detection more prepared in dealing with Advanced LIGO data.

\ack
% We are very grateful to Yanbei Chen and Josh Willis for many useful discussions and comments about our work.
We would like to thank Maurice H.P.M. van Putten for discussion and comments on the details of the paper. This research is supported in part by
National Natural Science Foundation of China (Grant No.~61440057, 61272087, 61363019 and 61073008), Beijing
Natural Science Foundation (Grant No.~4082016 and 4122039), the Sci-Tech Interdisciplinary Innovation and Cooperation Team Program of the Chinese Academy of Sciences,
the Specialized Research Fund for State Key Laboratories, and the Australian Research Council Discovery Grants and Future Fellowship programs. QC gratefully acknowledges the support of an International Postgraduate Research Scholarship funded by the Australian government. We thank Andre Fletcher for proofreading a draft of this manuscript.
%\end{acknowledgments}

\section*{References}
\bibliographystyle{unsrt}
% Create the reference section using BibTeX:
%\begin{thebibliography}{50},
\bibliography{references}
%\end{thebibliography}.
\end{document}